\documentclass[aps,prb,reprint,superscriptaddress,longbibliography]{revtex4-2}
\usepackage{graphicx}
\usepackage{latexsym}
\usepackage{amsmath}
\usepackage{amssymb}
\usepackage{amsfonts}
\usepackage{color}
\usepackage{bm}
\usepackage{verbatim}
\usepackage{hyperref}
\usepackage{empheq}
\usepackage[normalem]{ulem}
\graphicspath{{images/}}

\begin{document}

\title{
The uniqueness of the driven $\varphi_0$ Josephson junction: \\
when steps are not Shapiro
}
\author{K. Kulikov}
\affiliation{BLTP, JINR, Dubna, Moscow region, 141980, Russia}
\affiliation{Dubna State University, Dubna, Russia}

\author{J. Teki\' c}
\affiliation{"Vinča" Institute of Nuclear Sciences, Laboratory for Theoretical and Condensed Matter Physics - 020, University of Belgrade, PO Box 522, 11001 Belgrade, Serbia}

\author{E. Kovalenko}
\affiliation{Center for the Development of Digital Technologies, Krasnogorsk, Russia}

\author{M. Nashaat}
\affiliation{BLTP, JINR, Dubna, Moscow region, 141980, Russia}
\affiliation{Department of Physics, Faculty of Science, Cairo University, 12613, Giza, Egypt}

\author{T. A. Belgibayev}
\affiliation{BLTP, JINR, Dubna, Moscow region, 141980, Russia}

\author{Yu. M. Shukrinov}
\affiliation{BLTP, JINR, Dubna, Moscow region, 141980, Russia}
\affiliation{Dubna State University, Dubna, Russia}
\affiliation{Moscow Institute of Physics and Technology, Dolgoprudny 141700, Russia}

\date{\today}
\begin{abstract}
The  $\varphi_0$ superconductor-ferromagnet-superconductor Josephson junction exhibits unique locking phenomena under the external periodic signal when the magnetic component is taken into account. 
Contrary to the well-known Shapiro steps that come from the locking with the electric component, locking of the Josephson oscillations with the magnetic one results in the appearance of Buzdin steps in the current-voltage characteristic and a much more complex response of the system.
These steps possess distinctive properties that are indications of their unique origins and locking mechanisms.
The width of the Buzdin step oscillates with the amplitude of the magnetic component, nevertheless, it exhibits anomalies in the Bessel-like behavior.
In addition, we perform an analytical analysis that supports the numerical results and shows that the width of the Buzdin step represents a product of two Bessel functions.
Investigation of the effects that simultaneously appear in the magnetic subsystem reveals the presence of destructive interference and magnetization reorientation that accompany the appearance of Buzdin steps.

\end{abstract}
\maketitle

\section{Introduction}
\label{Intr}

Recent developments in superconducting spintronics and computation technologies sparked an intensive research of structures with the superconductor-ferromagnet (SF) interface~\cite{linder2015superconducting,eschrig2011spin,golubov2017controlling,mel2022superconducting}.
As two materials with radically different properties, they provide a unique paradigm where two antagonistic electron orders (zero resistance superconductivity versus the spin alignment in ferromagnets) under the right conditions can create a variety of new phenomena that offer a potential for the realization of high-speed and low consumption superconducting digital devices~\cite{ryazanov2012magnetic, cai2023superconductor,guarcello2020cryogenic}.
Among those structures, the one that is currently attracting a lot of attention from scientists and engineers is the superconductor-ferromagnet-superconductor (SFS) $\varphi_0$ Josephson junction (JJ).

The SFS $\varphi _0$ JJ, or simply $\varphi _0$ JJ, belongs to the special class of anomalous Josephson junctions with a non-centrosymmetric
ferromagnetic metal with broken inversion symmetry as a weak link~\cite{konschelle2009magnetic,buzdin2008direct,buzdin2005proximity,shukrinov2022anomalous}.
There, the Rashba-type spin-orbit coupling~\cite{amundsen2024colloquium} in the ferromagnetic layer leads to a phase shift proportional to the magnetic moment in the barrier.
As a result, the current-phase relation becomes $I=I_c\sin (\varphi -\varphi _0)$, where $I_c$ is the critical current, $\varphi$ is the superconducting phase difference, and $\varphi _0$ is the phase shift, which has been experimentally confirmed in different systems in recent years~\cite{szombati2016josephson, assouline2019spin,mayer2020gate}. 
The presence of {\it bidirectional} coupling between the magnetic moment of the barrier
and the superconducting phase difference opens up the possibility of controlling magnetism via supercurrent~\cite{konschelle2009magnetic} and vice versa, influencing Josephson current via magnetic moment~\cite{takahashi2007supercurrent} that has a huge scientific and technological potential~\cite{mel2022superconducting,ryazanov2012magnetic,rabinovich2019resistive,pal2018quantized,trahms2023diode,narita2022field}.
In contrast to regular Josephson junctions (with normal metal or insulator as a weak link) in the SFS junctions, the magnetic layer brings extra degrees of freedom and introduces the magnetic moment into the dynamics of a system~\cite{botha2023chaotic}.
This results in the appearance of a ferromagnetic resonance (FMR), which occurs without any external
radiation when the Josephson frequency becomes close to that of the FMR one~\cite{abdelmoneim2022locking,shukrinov2021anomalous,shukrinov2024buzdin,bobkova2022magnetoelectric}. 
The phenomenon is innate to the junctions with ferromagnetic interface and is absent in regular junctions. 
In the current-voltage ($IV$) characteristics, it manifests itself as a resonance branch over a
voltage interval that characterizes the width of the resonance, and as a sharp peak in the dependence
of the maximal value of the magnetic component on the bias current.

When a Josephson junction is submitted under external radiation, it exhibits locking phenomena manifested as the Shapiro steps (SS) in the current-voltage characteristics~\cite{shapiro1963josephson}.
However, due to the presence of a ferromagnetic weak link, an SFS $\varphi _0$ JJ  is also influenced by the magnetic component of external radiation.
The microwave magnetic field generates an additional magnetic precession with the microwave frequency, which results in a series of unusual phenomena ~\cite{shukrinov2024buzdin}. 
In addition to the Shapiro steps that result from the locking with the electric component of the radiation, two completely new types of steps are created by the periodic field of the magnetic component~\cite{shukrinov2024buzdin}.
In the case when only the magnetic component is present, as the magnetic moment precession becomes locked with the magnetic component, the Josephson oscillations also get locked due to their coupling with the ferromagnetic moment resulting in the appearance of novel steps called the Buzdin step (BS) since they were first predicted in Ref. \cite{konschelle2009magnetic}. 
When both components are present, the magnetic component drives the ferromagnetic moment, which further influences the superconducting current, while the electric component also interacts with the superconducting current. 
This leads to another type of step called the composite or chimera step (CS) due to its creation by two different mechanisms.
In the creation of both BS and CS, an important influence comes from the coupling of Josephson oscillations and ferromagnetic moment precession determined by the value of the spin-orbit parameter.
In addition to these steps, the presence of external radiation also leads to the appearance of resonance peaks whose properties depend on the periodic signal and the system parameters, opening up the possibility of controlling and manipulating the magnetic moment and resonance in hybrid SF systems~\cite{kulikov2024resonance}.

The dynamics of the regular JJs driven by an external periodic signal has been studied for decades, and numerous theoretical and experimental works have been done on how the amplitude and frequency of external radiation affect the system~\cite{benz1990fractional, sellier2004half,panghotra2020giant,tekic2016ac}.
However, most of those studies consider regular JJs under the influence of the electric component only, while the magnetic one is completely neglected. 
Any technological application of the SFS JJ 
demands a complete understanding of its behavior, tuneability, and control. 
These requirements motivated recent studies such as magnetization control of the critical current in 
a S-(S/F)-S superconducting switch~\cite{kammermeier2024magnetization}.
Nanoscale spin ordering and spin screening effects were investigated in tunnel ferromagnetic Josephson junctions where the nanoscale spin arrangement manifests itself directly in the magnetic dependence of the Josephson critical current~\cite{satariano2024nanoscale}.
In the search for the most suitable ferromagnetic materials for SFS junctions, the magnetic field dependence of the critical current is used as a common way to assess the quality of a Josephson junction~\cite{birge2024ferromagnetic}. 

In this work, we investigate the dynamics of the $\varphi_0$ Josephson junction under external radiation, focusing on the influence of the magnetic component. 
Unlike in the locking with the electric one, the system's response to the external periodic magnetic field is much more complex, characterized by the appearance of not just Buzdin steps but also the phenomena that simultaneously occur in the magnetic subsystem.
One may easily assume that the Buzdin steps are just another type of Shapiro steps, however, they behave in a unique way due to their different physical origins.
In the amplitude dependence, they exhibit anomalies in the Bessel-like behavior typical for Shapiro steps.
In contrast to the case of locking with the electric component, locking with the magnetic one does not influence the critical current (zero step).
Furthermore, we perform an analytical investigation that supports the numerical results and provides a deeper insight into physics.
To get a complete picture of the system dynamics we also analyze the behavior of the magnetic subsystem, which reveals an interesting phenomenon that appears in the $\varphi _0$ JJ under the magnetic component of external radiation: the locking accompanied by the destructive interference and magnetization reorientation.

The paper is organized as follows.
The model is introduced in Sec. I\ref{Mod}. 
The peculiarities of the Buzdin steps are studied in Sec. \ref{BSvsSS}.
The analytical calculation of the Buzdin step width is given in Sec. \ref{anal}.
Sec. \ref{Antires} discusses the phenomena that simultaneously appear in the magnetic subsystem. 
Finally, Sec. \ref{Con} concludes the paper.

\section{Model}
\label{Mod}

\begin{figure}
\centering
         \includegraphics[width=0.9\linewidth]{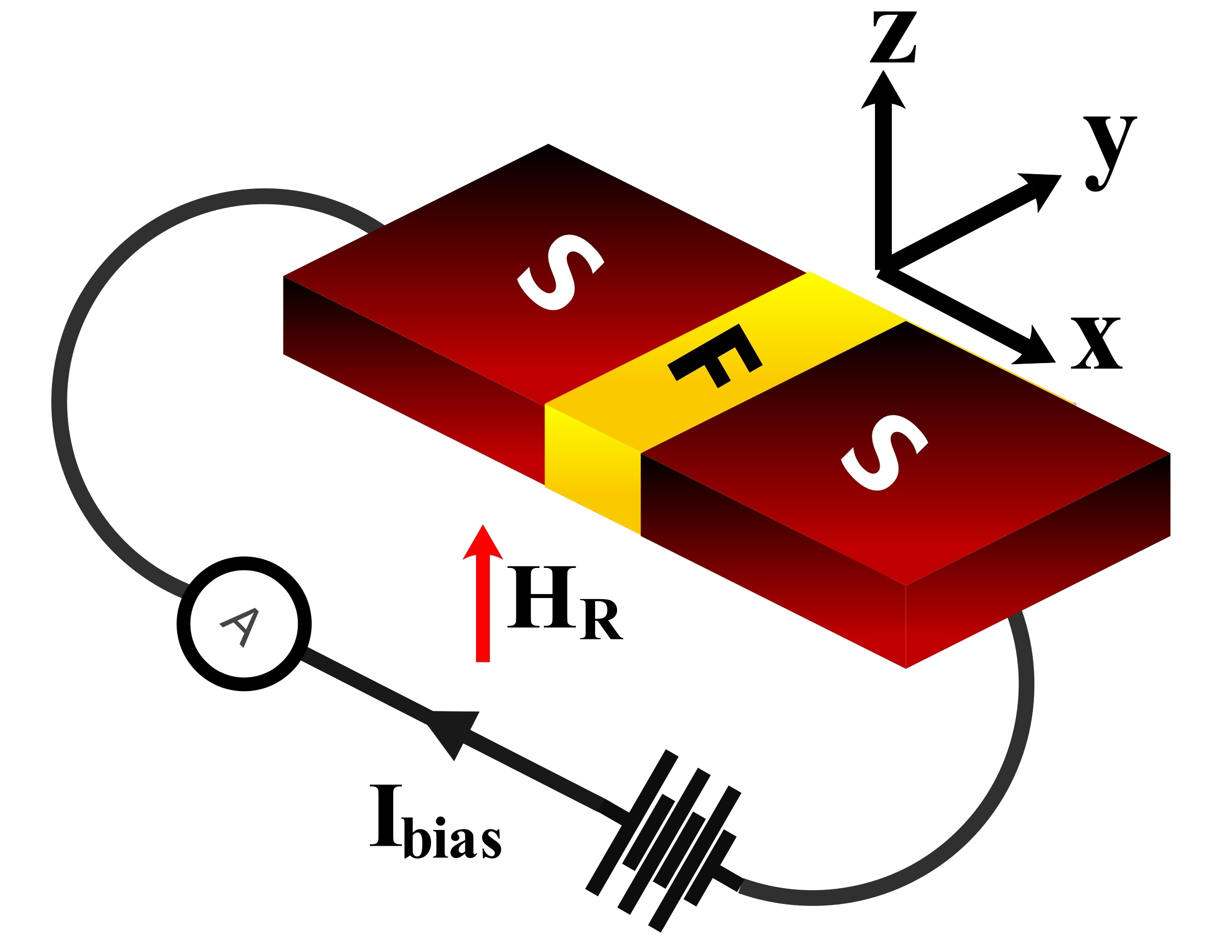}
	\caption{The geometry for current biased SFS $\varphi _0$ Josephson junction, the bias current flow in the $x$-direction, the ferromagnetic easy axis, and the ac magnetic field are in the $z$-direction.
 }
\label{Fig1}
\end{figure}

We consider the current biased $\varphi _0$ SFS JJ with the geometry presented in Fig.~\ref{Fig1} under external radiation, where, for the study of Buzdin steps, only the magnetic component is taken into account.
Here, the ferromagnet easy axis and the gradient of the spin-orbit potential are directed along the $z$ axis. 
Accordingly, the additional phase shift $\varphi_0$ is proportional to the $y$ component of the magnetic moment $\textbf{M}$ \cite{konschelle2009magnetic,shukrinov2024buzdin}.

Josephson junctions with the ferromagnetic weak link are characterized by the coupling between the superconducting phase difference across the junction $\varphi$ and the magnetization $\mathbf{M}$ of the ferromagnetic (F) layer.
Thus the dynamics of $\varphi _0$ SFS JJ can be described by the system of equations obtained from the Landau-Lifshitz-Gilbert (LLG) equation~\cite{lif91course}, the Josephson relation for the phase difference and voltage, and the equation for the biased current of the resistively and capacitively shunted junction (RCSJ) model~\cite{tinkham2004introduction}.

In the dimensionless form, the total system of LLG-Josephson equations reads as \cite{shukrinov2024buzdin}:
\begin{eqnarray} \label{LLG_components}
	\dfrac{dm_{x}}{dt} &=& \frac{\omega_{F}}{(1+\alpha^{2})} \bigg[ h_y \left(m_z-\alpha  m_x m_y\right)\nonumber \\
	&& - h_z \left(\alpha  m_x m_z+m_y\right)+\alpha  h_x \left(m_y^2+m_z^2\right) \bigg] , \nonumber \\
	\dfrac{dm_{y}}{dt} &=&\frac{\omega_{F}}{(1+\alpha^{2})} \bigg[ -h_x \left(\alpha  m_x m_y+m_z\right)\nonumber \\
	&& + h_z \left(m_x-\alpha  m_y m_z\right)+\alpha  h_y \left(m_x^2+m_z^2\right) \bigg] , \nonumber \\
	\dfrac{dm_{z}}{dt} &=&\frac{\omega_{F}}{(1+\alpha^{2}) } \bigg[  \alpha  h_z \left(m_x^2+m_y^2\right) \nonumber\\
	&& - h_y \left(m_x+\alpha  m_y m_z\right)+h_x \left(m_y-\alpha  m_x m_z\right) \bigg] ,\nonumber \\
    \dot{V} &&= \left[ I-V(t)+r\dot{m}_{y}-\sin (\varphi -rm_{y})\right]/\beta_c, \nonumber \\
	\dot{\varphi} &&= V(t),
    \label{LLGJ}	
\end{eqnarray}
with the effective field, which is given by:
\begin{eqnarray}
h_{x} &=& 0,  \nonumber \\
h_{y} &=& Gr \sin(\varphi - r m_{y}) , \\ 
h_{z} &=& m_{z}  + \dfrac{h_r}{\omega_{F}}\sin(\omega_R t) \nonumber
\label{Effective_Field_components}
\end{eqnarray}
where the magnetization components are normalized to the saturation value $M_{s}$, $\alpha$ is the Gilbert damping, the time is normalized to $(\omega _c)^{-1}$, where $\omega _c=2eI_cR/\hbar $ is the characteristic frequency of the junction. The ratio of the Josephson to magnetic anisotropy energy is given by ${G=E_J/E_{an}}$. 
The relative strength of spin-orbit coupling is characterized
by the Rashba type parameter $r$~\cite{konschelle2009magnetic}.
The variables $\omega _F$ and $\omega_{R}$ are the normalized frequencies for the ferromagnetic resonance and external radiation, respectively, and scale as $\omega _c$. The amplitude of the magnetic component is normalized so that $h_R=\frac{\gamma }{\omega _c}H_R$, $\gamma$ is the gyromagnetic ratio. 
In Eqs. (\ref{LLGJ}), $\beta_{c}=2eI_{c}CR^{2}/\hbar$ is the McCumber parameter (here, we consider $\beta_{c}=25$), $I_{c}$, $C$, $R$ are the Josephson critical current, the resistance, and the capacitance, respectively. The normalized bias current $I$ scales as $I_{c}$, while the normalized $V$ across the junction scales as $I_{c}R$, respectively.
According to the above normalizations $V=\omega _J$. 

This system of equations was integrated numerically, where for each step of the external bias current, $m_i(t) (i=x,y,z)$, $V(t)$, and $\varphi(t)$ were calculated. 
Then, using the averaging procedure from Refs.~\cite{shukrinov2007investigation,buckel2008superconductivity}, the average voltage and hence the $IV$-characteristic are determined. 
The Buzdin steps are analyzed for different system parameters.

\section{Buzdin versus Shapiro steps}
\label{BSvsSS}

It is well known that the width of Shapiro steps exhibits Bessel-like oscillations with the amplitude of external radiation~\cite{benz1990fractional,tekic2016ac}.
These oscillations are correlated so that the first harmonic maxima correspond to the critical current (zero step) minima and vice versa.
However unlike the locking with the electric component, locking with the magnetic one results in steps with very different properties.

\begin{figure}[tbh]
 \centering
  \includegraphics[width=0.7\linewidth]{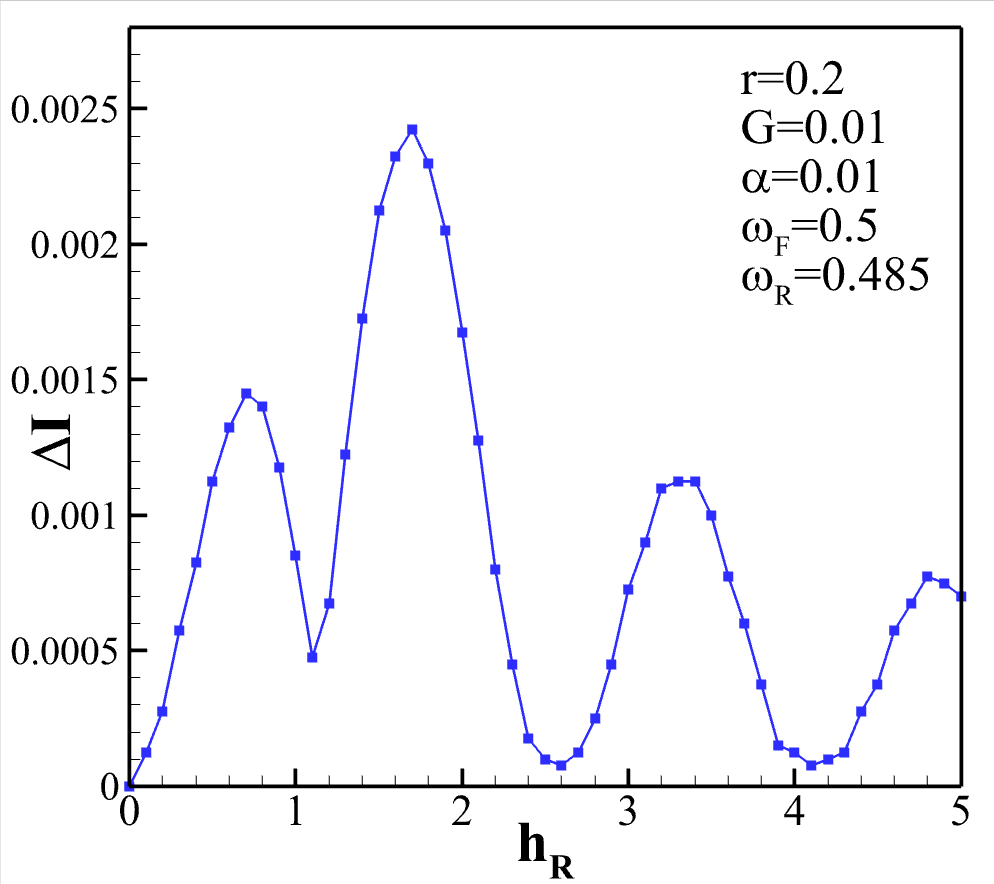}
\caption{ The size of the first harmonics BS $\Delta I$ 
as a function of the amplitude of the magnetic component of the microwave field $h_{R}$ for $r=0.2$, $G=0.01, \alpha =0.01, \omega _F=0.5, \omega _R=0.485$, and $\beta _c=2$. 
}
\label{Fig2}
\end{figure}

We consider the case when the frequency of external radiation $\omega _R$ is close to the FMR frequency 
$\omega _F$, and investigate the BS that appears at 
$\omega _J=\omega _R\approx \omega_F$.
The amplitude dependence of the BS is presented in Fig. \ref{Fig2}.
The width of BS oscillates with the magnetic component $h_R$ of external radiation; nevertheless, it exhibits anomalies.
The Bessel-like oscillations, i.e., the standard behavior typical for Shapiro steps, are absent.
Instead, we can see that the first maximum is lower than the second one.
Later in Sec. \ref{anal} by performing analytical analysis we will show that this particular amplitude dependence comes as a result of the product of two Bessel functions.
During our investigation, we also observed that, unlike in the case of locking with the electric component, locking with the magnetic one did not affect the critical current, which remained constant.

\section{Analytical analysis}
\label{anal}

To gain an insight into the physics behind the Buzdin steps we provide an analytical analysis of their behavior (details can be found in the Appendix).

The dynamics of the JJ is described within the RCSJ model by the following equation:
\begin{eqnarray}
\beta_c \Ddot{\varphi} + \dot{\varphi} +\sin{(\varphi -r m_y)}=I+r\dot{m}_y.
\label{phi}
\end{eqnarray}
Since all the numerical calculations have been done in the current bias regime, we are using the “Harmonic perturbation theory” (HPT) \cite{fominov2022, fominov2024, tompson1973} in the large-capacitance limit near Ohm’s law where $\omega_J = I$ (accurately $\beta_c I\gg 1$) to calculate the Josephson phase and dc current corrections. 
We look for the solution in the form:

\begin{eqnarray}
\varphi(t) &=& \phi_0 + \omega_J t +\sum_{n} \left[ a_n \cos(n\omega_J t ) + b_n \sin(n\omega_J t )\right].
\label{phi-harmonic}
\end{eqnarray}
The phase $\varphi$ and dc current $I$ can be written as perturbative series:

\begin{eqnarray}
 \varphi(t)&=& \phi_0 +\omega_J t+\sum_{k=1} \varphi^{(k)}  \hspace{1cm} I = \omega_J +\sum_{k=1} \overline{j}^{(k)} \nonumber \\
 a_n&=&\sum_{k=1} a_n^{(k)}   \hspace{2cm}  b_n=\sum_{k=1} b_n^{(k)}
\label{current-expansion}
\end{eqnarray}
where $ \phi_0$ is initial phase, $\varphi^{(k)}$ is contributions to the phase across the junction, $\overline{j}^{(k)}$ is the average current in the $k$-th order of the perturbation theory, and $a_n$ and $b_n$ are the amplitudes of the $n$-th harmonic.
In Eq. (\ref{phi-harmonic}), the linearly growing term is exactly equal to $\omega_J t$, while the remaining part oscillates with frequencies that are multiples of $\omega_J$. Note that we do not expand $\cos(n\omega_Jt)$ and $\sin(n\omega_Jt)$ into series. In this method, corrections to the average current $\overline{j}^{(k)}$ appear only due to constant (nonoscillating) terms generated by products of trigonometrical functions.

Substituting equation (\ref{phi-harmonic}) into (\ref{phi}), expanding the resulting equation into the Fourier series, and taking into account only linear and first-order terms, we obtain:

\begin{eqnarray}
-\beta_c \omega_J^2 \Bigg[ a_1  \cos(\omega_J t ) + b_1 \sin(\omega_J t )\Bigg] - \overline{j}^{(1)}  = \nonumber \\ 
= - \sin( \phi_0 +\omega_J t )+ r m_y(t)\cos( \phi_0 +\omega_J t)  +r\dot{m}_y.
\label{phin1}
\end{eqnarray}
where $a_1$ and $b_1$ will give us the amplitude of $\varphi(t)$, and $\overline{j}^{(1)}$ represents the constant correction to the current, i.a., the current step. 
To determine the width of the BS we need to calculate $\overline{j}^{(1)}$ using Eq. (\ref{phin1}).
The average from the first and last terms gives zero (see Appendix), and the term that remains is $\left< r m_y(t)\cos( \phi_0 +\omega_Jt)\right>$. 
This term will give an additional average superconducting current $\overline{j}^{(1)}=\left<r m_y(t)\cos( \phi_0 +\omega_Jt)\right>$, in other words, the value of the step width. 

First, we need to find the projection of the magnetization on the $y$-axis $m_y(t)$. This can be done in the ''weak coupling`` regime $G \ll 1$ when the Josephson energy is small in comparison with the magnetic one, and the magnetic moment precesses around the z-axis. 
If the other components verify $(m_x, m_y) \ll 1$, then the system of LLG equations can be linearized as:
\begin{eqnarray}
	\dot{m_{x}} + \omega_{F} \left(1 + \dfrac{h_R}{\omega_{F}} \sin(\omega_R t)\right) m_y &=& \omega_{F} rG \sin( \phi_0 +\omega_Jt) , \nonumber \\
	\dot{m_{y}} - \omega_{F} \left(1 + \dfrac{h_R}{\omega_{F}} \sin(\omega_R t)\right) m_x  &=& 0, 
 \label{LLG_components}
\end{eqnarray}
We note that in our calculations $\alpha \ll 1 \Rightarrow \frac{\omega_{F}}{(1+\alpha^{2})} \approx  \omega_{F}$, and since $r < 1$ then $rm_y$ is much smaller than $m_y$, therefore not included into consideration. The inhomogeneous solution for $m_y$ component, in this case, is given by: 

\begin{widetext}
\begin{eqnarray}\label{m_y}
	  m_y &=& \dfrac{rG\omega_{F} }{2} Re\left(  \sum^{\infty}_{n=-\infty}i^{n-1}J_n\left(\dfrac{h_R}{\omega_{R}}\right)\left[\sum^{\infty}_{q=-\infty}i^{q-n}(-1)^{q-n}J_{q-n}\left(\frac{h_R}{\omega_{R}}\right)\dfrac{e^{i( q\omega_R t - \omega_J t- \phi_0)}}{ n\omega_R -\omega_{F} - \omega_J } \right. \right.\nonumber \\ 
	  &-& \left. \left.\sum^{\infty}_{q=-\infty}i^{q-n}(-1)^{q-n}J_{q-n}\left(\frac{h_R}{\omega_{R}}\right)\dfrac{e^{i(q\omega_R t +  \omega_J t+\phi_0)}}{n\omega_R -\omega_{F} +  \omega_J} \right] \right).\nonumber \\
\end{eqnarray}
\end{widetext}
This solution can be substituted into the Josephson equation to calculate $r m_y(t)\cos(\phi_0 + \omega_Jt)$:
\begin{widetext}
\begin{eqnarray}\label{cos_term}
	  r m_y(t)\cos(\omega_Jt+\phi_0) &=& \dfrac{r^2G\omega_{F} }{4} Re\left(  \sum^{\infty}_{n=-\infty}i^{n-1}J_n\left(\dfrac{h_R}{\omega_{R}}\right)\left[ \sum^{\infty}_{q=-\infty}i^{q-n}(-1)^{q-n}J_{q-n}\left(\frac{h_R}{\omega_{R}}\right)\dfrac{e^{i( q\omega_R t - 2\omega_J t-2\phi_0)}}{ n\omega_R -\omega_{F} - \omega_J } \nonumber \right. \right. \\ 
	  &-& \left. \left. \sum^{\infty}_{q=-\infty}i^{q-n}(-1)^{q-n}J_{q-n}\left(\frac{h_R}{\omega_{R}}\right)\dfrac{e^{i(q\omega_R t + 2 \omega_J t+2\phi_0)}}{n\omega_R -\omega_{F} +  \omega_J} \right] \right).\nonumber \\
\end{eqnarray}
\end{widetext}

The Buzdin step appears when $q\omega_R - 2\omega_J=0$, which leads to
the condition for BS:
\begin{eqnarray}
\omega_J=\dfrac{q\omega_R}{2},
\label{BScond}
\end{eqnarray}
where $q=1,2,...$ corresponds to the first, second, etc. harmonics.
Further, we will determine the size of the BS at $\omega_J=\omega_R$.
Taking into account that we need only the largest terms in Eq. (\ref{cos_term}), we cionsider two cases: when the driving frequency $\omega_R$ is in the FMR region and when it is near $\omega_{F}/2$.


When the driving frequency is close to the FMR one, i.e., $\omega_J=\omega_R \approx \omega_F$, in which case $\omega_J -\omega_{F} \ll 1$, we have the following conditions for $q$ and $n$:
\begin{eqnarray}
	  q - 2 &=& 0, \hspace{1cm} n - 1 = 1, \nonumber \\
	  q + 2 &=& 0, \hspace{1cm} n + 1 = 1. \nonumber
\end{eqnarray}
\label{nk}
Thus, the pairs of $q$ and $n$ for the first and second terms are the following:

\begin{eqnarray}
	  n &=& 2, \hspace{1cm} q = 2, \nonumber \\
	  n &=& 0, \hspace{1cm} q = -2, \nonumber
   \label{nk12}
\end{eqnarray}

Substituting those into (\ref{cos_term}) and taking into account that $\dfrac{e^{i2\phi_0}+e^{ -i2\phi_0}}{2i}=\sin(2\phi_0)$ one can get:
\begin{widetext}
\begin{eqnarray}\label{cos_term_4}
	  \overline{j}^{(1)}=\overline{ rm_y(t)\cos(\omega_Rt+\phi_0)} &=& \dfrac{r^2G\omega_{F} }{2(\omega_R -\omega_{F})} J_2\left(\dfrac{h_R}{\omega_{R}}\right)J_0\left(\dfrac{h_R}{\omega_{R}}\right)\sin(2\phi_0).\nonumber \\
\end{eqnarray}
\end{widetext}
Then, the BS width 
can be calculated as:
\begin{widetext}
\begin{eqnarray}\label{Dj}
\Delta j^{(1)}=\max \overline{j}^{(1)} -
\min \overline{j}^{(1)} = \frac{r^2G\omega _F}{(\omega_R -\omega_{F})}J_2\left( \dfrac{h_R}{\omega_{R}}\right)J_0\left(\dfrac{h_R}{\omega_{R}}\right).\nonumber \\
\end{eqnarray}
\end{widetext}
The presented analytical analisis shows that the Buzdin steps appears due to an additional phase term $(rm_y)$ in the supercurrent (see Eq. \ref{phi}) and that the additional quasiparticle term
$(r \dot{m}_y)$ does not contribute to the size of the step in the first order of perturbation theory.

\begin{figure}[tbh]
 \centering
  \includegraphics[width=0.6\linewidth]{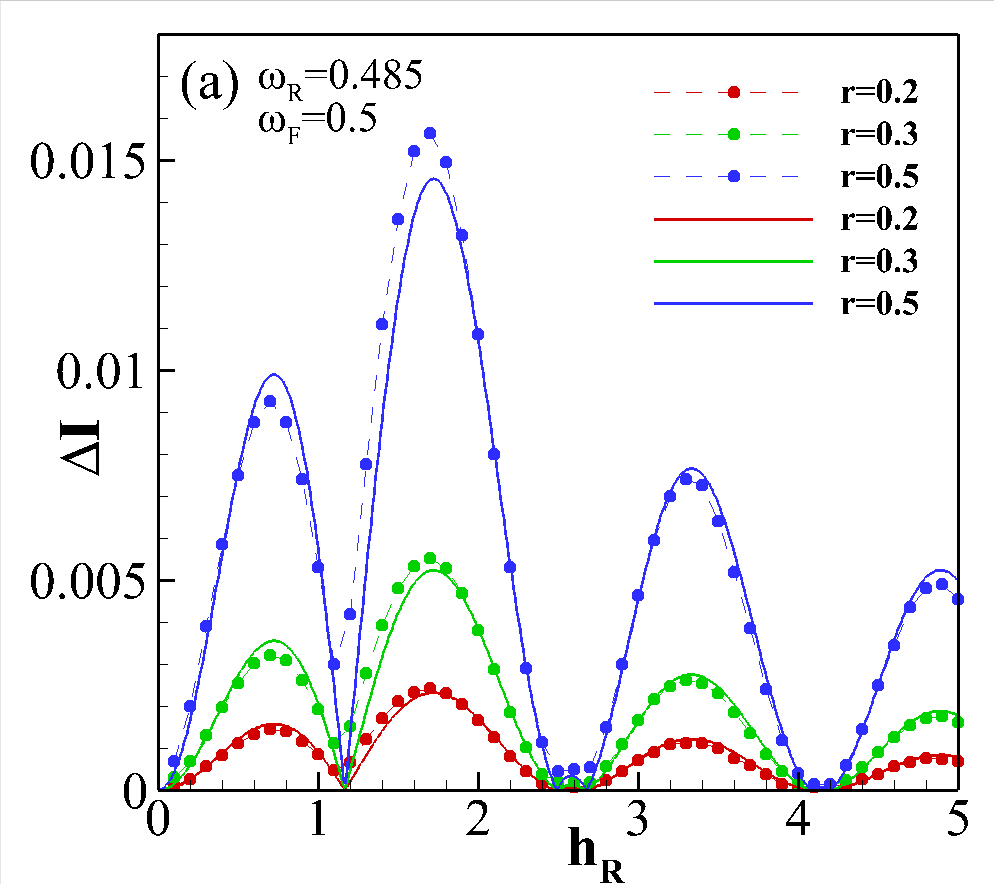}
  \includegraphics[width=0.6\linewidth]{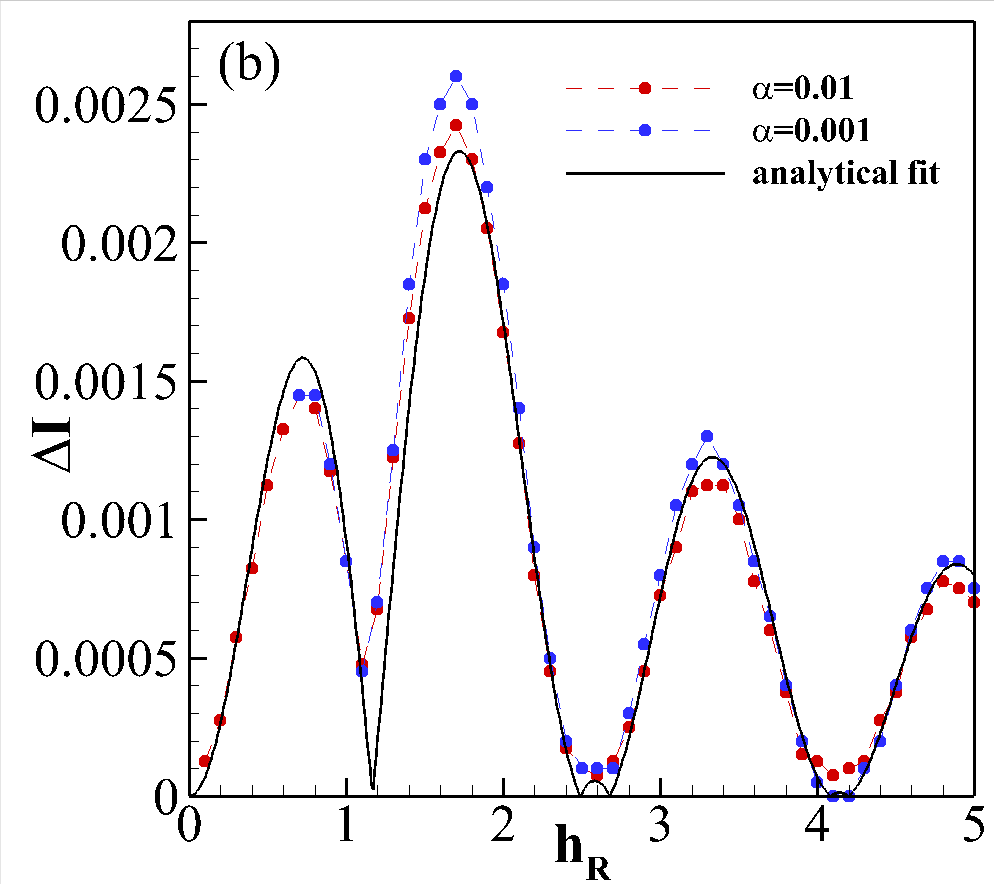}
\caption{The comparison between the numerical and analytical results for the size of Buzdin step $\Delta I$ as a function of the amplitude of the magnetic component $h_R$ at $\omega_J=\omega_R\approx \omega _F$ for different values of $r$ and $\alpha =0.01$ in (a), and different values of $\alpha$ and $r=0.2$ in (b). Dot (solid) lines correspond to the numerical (analytical) results. 
The rest of the parameters are the same as in Fig. \ref{Fig2}
}
\label{Fig3}
\end{figure}

The comparison between the numerical and analytical results for different values of $r$ and $\alpha$ is presented in Fig. \ref{Fig3}(a) and (b), respectively.
The spin-orbit coupling determines the size of BS, which gets larger as $r$ increases in Fig. \ref{Fig3} (a), but does not qualitatively change its amplitude dependence.
The effect of Gilbert dumping can be seen in Fig. \ref{Fig3} (b). The analytical result was obtained in the limit $\alpha \ll 1$, and our numerical studies, performed for $\alpha =0.01$, agree very well with the analytical one.
The changes of $\alpha$ do not affect the form of oscillation and have a relatively small influence on the BS, where the decrease of $\alpha$ slightly increases its size.

\begin{figure}[tbh]
 \centering
  \includegraphics[width=0.6\linewidth]{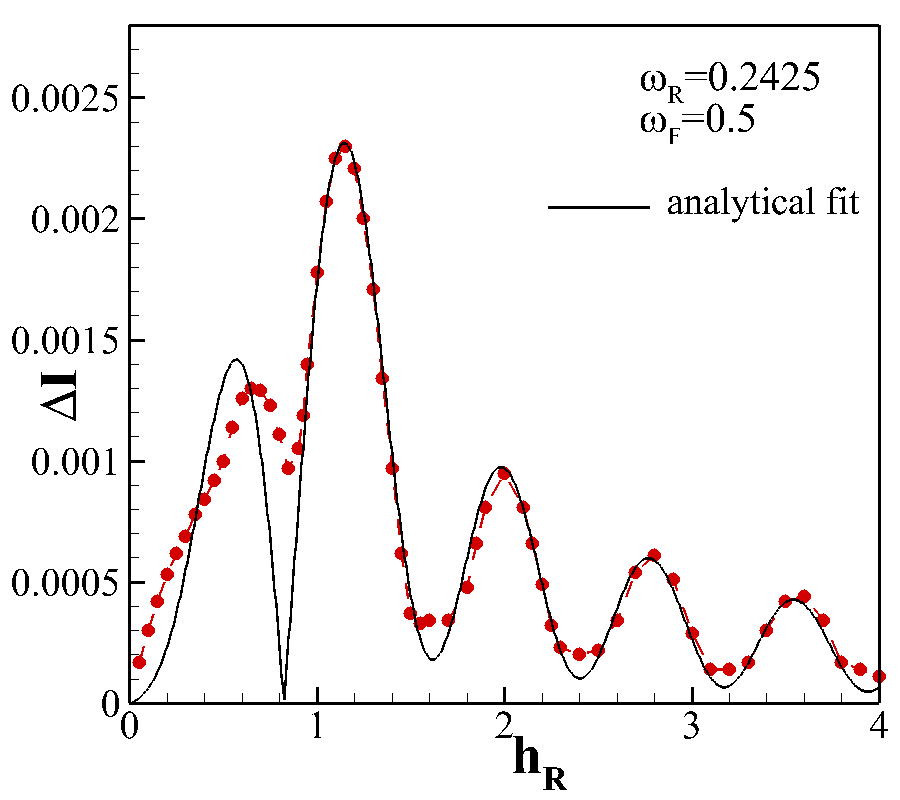}
\caption{The comparison between the numerical and analytical results for the size Buzdin step $\Delta I$ at $\omega_J=\omega_R\approx \omega _F/2$ as a function of the amplitude of the magnetic component $h_R$ for $\omega_F=0.5$ and $\omega_R=0.2425$.
Dot (solid) lines correspond to the numerical (analytical) results.
The rest of the parameters are the same as in Fig. \ref{Fig2}
}
\label{Fig4}
\end{figure}

To investigate BS when $\omega _R$ is out of the FMR region we will consider the case when $\omega_J=\omega_R \approx \omega_F/2$.
Following the same procedure as in the derivation of
Eq. (\ref{Dj}), we obtain the BS width given as (see Appendix):
\begin{eqnarray}
\Delta j^{(1)}  = 
\dfrac{r^2G\omega_{F} }{2\omega_R -\omega_{F}}J_3\left(\dfrac{H_r}{\omega_{R}}\right) J_{1}\left(\frac{H_r}{\omega_{R}}\right).
\label{DjwF/2}
\end{eqnarray}
The numerical and analytical results are compared in Fig. \ref{Fig4}.
Unlike the results in Fig. \ref{Fig3}, here at low values of $h_R$, the amplitude dependence of BS does not correspond well to the product of two Bessel functions and exhibits anomalies.

Another property that distinguishes the SFS JJ driven by the magnetic component from the systems driven by the electric component of external radiation is the critical current (zero step) behavior.
The critical current is uncorrelated with BS and remains constant with the changes in $h_R$. 
In this case, $\omega_J=0$ and an average over time from Eq.(\ref{cos_term}) will become zero as it has been shown in Appendix \ref{ZVS}. 
Therefore, there is no contribution from the magnetic subsystem to the current value of the zero-voltage state.
We must stress that this is not a general rule for SFS JJs, as in the SFS JJ on the topological insulator, the critical current is modulated by the magnetic subsystem~\cite{nashaat2019electrical}.

\vspace{2 cm}

\section{Destructive interference}
\label{Antires}

Until now, we were focused on the BS and the behavior of the superconducting subsystem.
In this section, we consider the ferromagnetic one and investigate how the locking of the junction with the external magnetic component influences its dynamics.
During our investigations, we observed that 
as the junction gets locked, the appearance of BS is accompanied by sudden changes in the magnetisation dynamics.
 
\begin{figure}[tbh]
	\centering
	\includegraphics[width=0.95\linewidth]{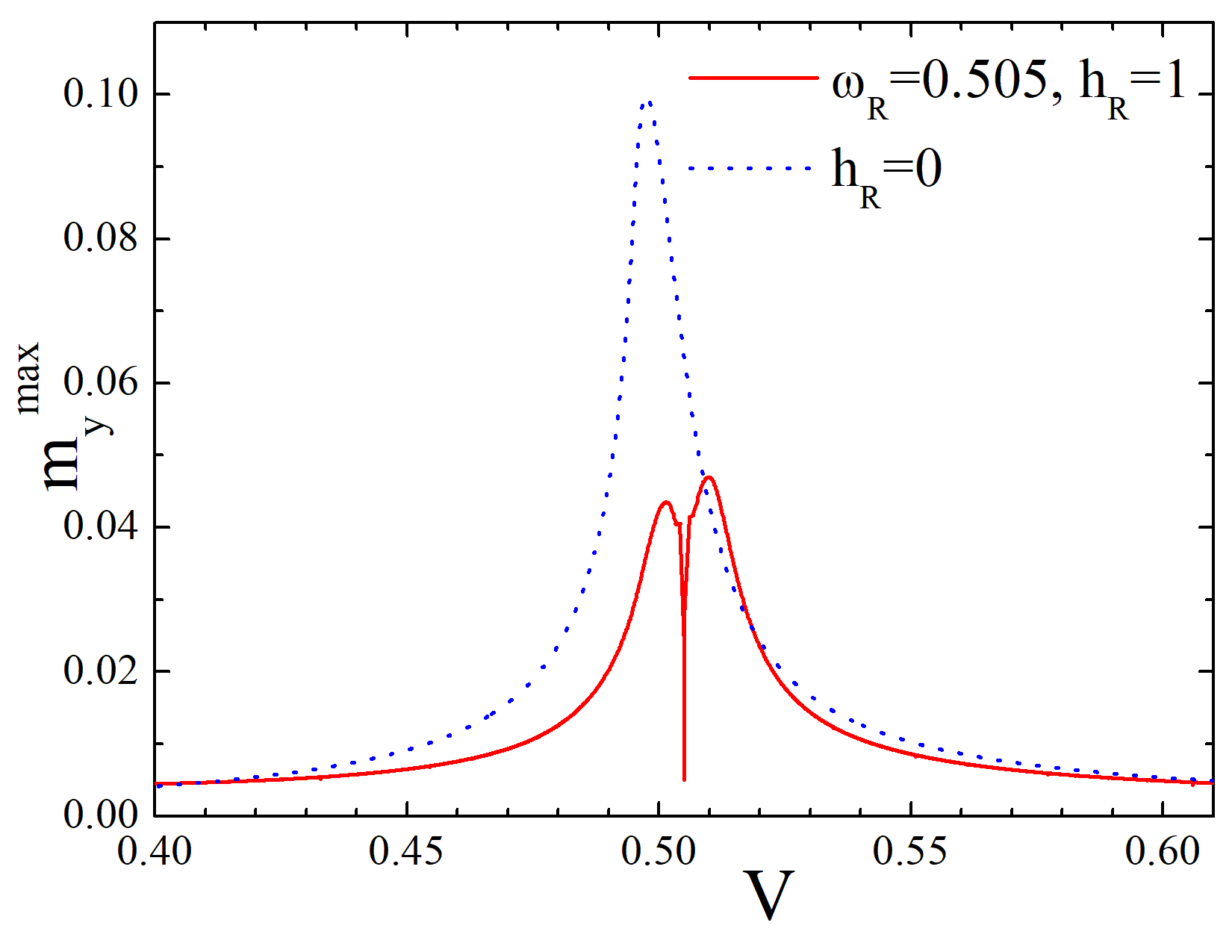}
	\caption{$m_{y}^{max}$ as a function of voltage $V$ in the presence (solid line) and in the absence (dashed line) of the external ac magnetic field at $G=0.01, \alpha=0.01, r=0.2, \omega_F=0.5$.}
	\label{Fig5}
\end{figure}

The sudden drop in the amplitude of magnetization precession is presented in Fig.\ref{Fig5}, where the maximum value of the $m_y$ component is plotted as a function of voltage $V$ with (solid line) and without (dashed line) ac magnetic field, respectively. 
While in the absence of external radiation $(h_R=0)$, $m_{y}^{max}$ exhibits a resonance peak, in its presence when the external frequency is in the FMR region, i.e., $\omega _r\approx\omega_F$, the amplitude of magnetic precession drops significantly at $\omega_J=\omega_R$.
In our system, this drop is caused by destructive interference between the external ac magnetic field and the Josephson oscillations. 

\begin{figure}[tbh]
	\centering
	\includegraphics[width=0.9\linewidth]{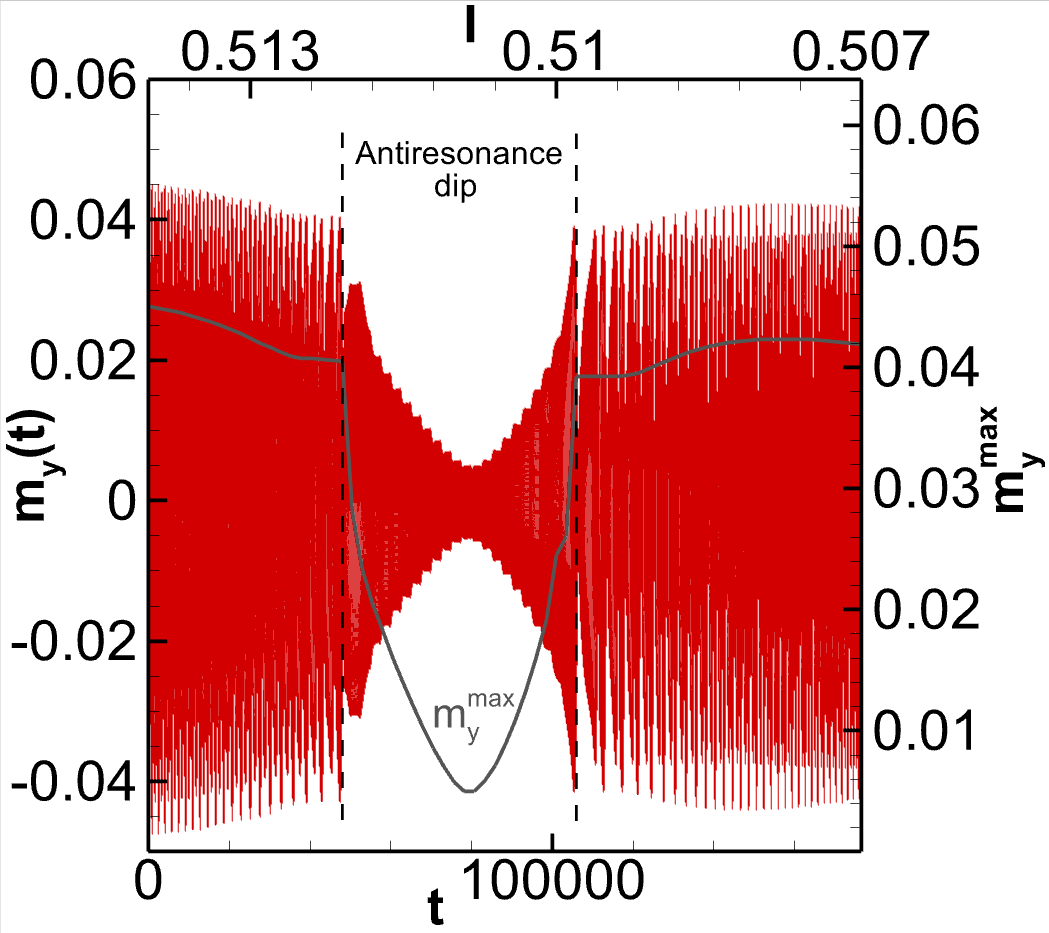}
	\caption{Time dependence of $m_y$-component (red line) together with $m_{y}^{max}$ as a function of bias current $I$ (grey line) at $G=0.01, \alpha=0.01, r=0.2, \omega_F=0.5, h_R=1, \omega_R=0.505$. }
	\label{Fig6}
\end{figure}

\begin{figure}[tbh]
	\centering
	\includegraphics[width=0.95\linewidth]{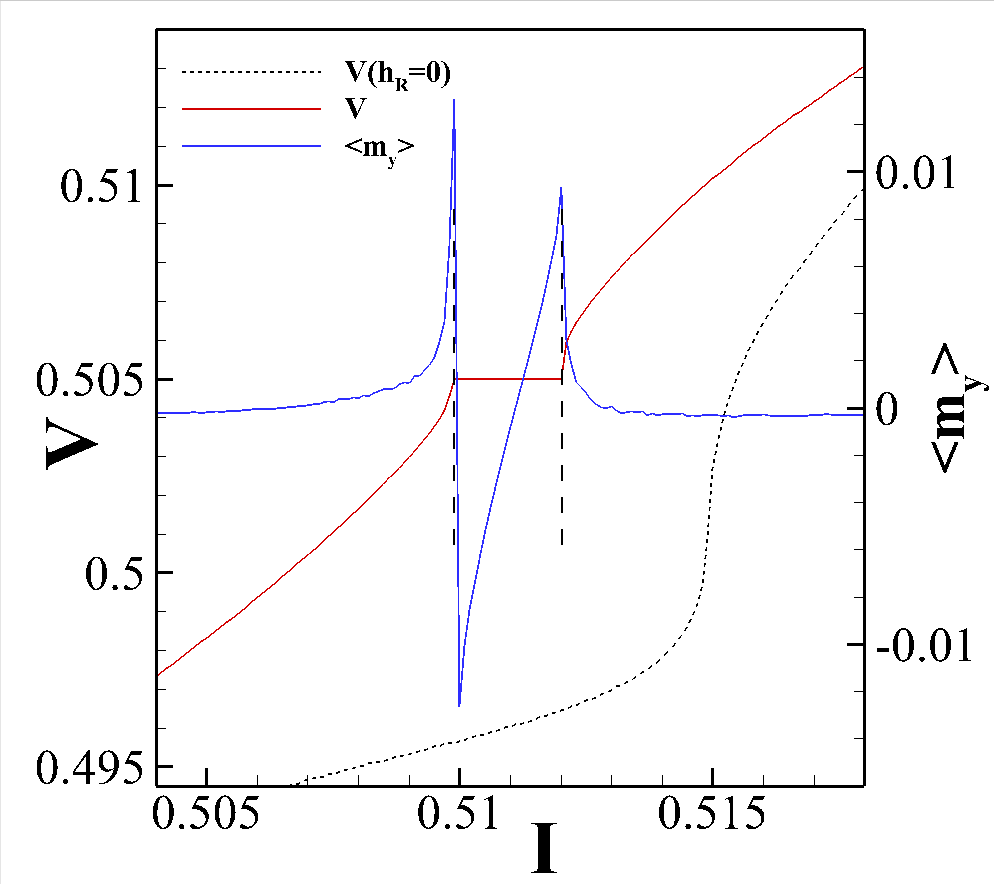}
	\caption{The current voltage-characteristics $V(I)$ and  $\left<m_y\right>$ as functions of $I$.
    The dotted curve corresponds to the case without external radiation. The rest of the parameters are as in Fig.\ref{Fig6}
}
	\label{Fig7}
\end{figure}

The dynamics of the magnetic subsystem can be analytically described by (\ref{m_y}), which is a summation of an infinite number of oscillating terms. The condition for destructive interference can be obtained by substituting the locking condition $\omega_J=\omega_R$ into Eq. (\ref{m_y}):  
\begin{widetext}
\begin{eqnarray}\label{phy_m_int_antir}
	  m_y &=&\dfrac{rG\omega_{F} }{2(\omega_R -\omega_{F})} J_1\left(\dfrac{h_R}{\omega_{R}}\right)\left(J_0\left(\dfrac{h_R}{\omega_{R}}\right)-J_2\left(\dfrac{h_R}{\omega_{R}}\right)\right)\cos(\phi_0)+\nonumber \\
	  &+& \dfrac{rG\omega_{F} }{2} Re\left(  \sum^{\infty}_{n\neq 2, n=-\infty}i^{n-1}J_n\left(\dfrac{h_R}{\omega_{R}}\right)\sum^{\infty}_{q\neq 1, q=-{\infty}}i^{q-n}(-1)^{q-n}J_{q-n}\left(\frac{h_R}{\omega_{R}}\right)\dfrac{e^{i( (q - 1)\omega_R t-\phi_0)}}{ (n-1)\omega_R -\omega_{F} } - \right. \nonumber \\ 
	  &-& \left. \sum^{\infty}_{n\neq 0, n=-\infty}i^{n-1}J_n\left(\dfrac{h_R}{\omega_{R}}\right)\sum^{\infty}_{q\neq -1, k=-\infty}i^{q-n}(-1)^{q-n}J_{q-n}\left(\frac{h_R}{\omega_{R}}\right)\dfrac{e^{i((q + 1)\omega_R t+\phi_0)}}{(n+1)\omega_R -\omega_{F} }  \right).
\end{eqnarray}
\end{widetext}
We notice in Eq.~\ref{phy_m_int_antir} that the main contribution to the amplitude of precession in the resonance condition comes from the terms with denominators $<<1$: $(n-1)\omega_R -\omega_{F}$ and $(n+1)\omega_R -\omega_{F}$ for $n=2$ and $n=0$, respectively, but two terms with $q=1$ and $q=-1$ among those became time independent and do not contribute to the amplitude anymore. 
Therefore, two external influences in the magnetic subsystem, the ac magnetic field and the Josephson oscillations, mutually cancel each other at $\omega_J=\omega_R$, leading to an abrupt drop in the amplitude of magnetization precession.
Fig.\ref{Fig6} shows the correspondence between the dip area in the $m_{y}^{max}(I)$ dependence and the area of significant reduction of the $m_y$ oscillation amplitude. 
As the system is locked, the average superconducting current also creates an additional constant contribution to the oscillating effective magnetic field that leads to the magnetization reorientation, as can be seen in Fig. \ref{Fig7}.

\section{Conclusion}
\label{Con}

Just as Shapiro steps are associated with the ac driven conventional Josephson junctions, the Josephson junctions with the ferromagnetic interface are characterized by the appearance of Buzdin steps when exposed to the magnetic component of the external radiation.
In this work, we presented the effects that the magnetic component of external radiation had on the $\varphi_0$ JJ, focusing not just on the response of the junction but also on the response of the magnetic subsystem.
Due to their different origin, the Buzdin steps exhibit unique properties.
In the amplitude dependence, the width of the Buzdin step demonstrates anomalous oscillations, and the Bessel-like behavior, typical for the Shapiro step, is absent.
Analytical analysis further confirms this behavior and shows that these anomalous oscillations result from the product of two Bessel functions.
Unlike in junctions driven by the electric component, in the $\varphi_0$ JJ, the magnetic component does not influence the critical current, which remains constant and uncorrelated with the behavior of the Buzdin steps.
Investigation of the magnetic subsystem shows that locking in the superconducting subsystem is accompanied by a sudden drop in the amplitude of magnetization precession and the magnetization reorientation in the magnetic subsystem.
  
An important property, not addressed in this work, is the frequency dependence of Buzdin steps.
Unlike in the case of Shapiro steps, where amplitude and frequency dependence are correlated so that their size oscillates with the frequency of the electric component, in the case of Buzdin steps, this correlation is absent. 
Our preliminary investigation shows that Buzdin steps exhibit huge resonant peaks at the harmonic and subharmonic values of the ferromagnetic resonance frequency because of the underlying mechanism of the ferromagnetic resonance that is present in SFS JJ.
The role of ferromagnetic resonance and the frequency dependence of Buzdin steps will be published separately.

Another interesting problem for future investigation comes from the fact that the presented results were obtained for the current biased $\varphi_0$ JJ and in one particular geometry, which leaves the question of how the other systems, with different biases or geometries, would respond to the driving of the magnetic component.
In comparison to the Shapiro steps, the Buzdin steps have a significantly smaller size, and for the purpose of technological application of SFS Josephson junctions, it would be interesting to examine whether it is possible to create large Buzdin steps comparable to Shapiro ones.

\begin{acknowledgments}
Numerical simulations were funded by the Russian Science Foundation Project No. 22-71-10022.
J. T. would like to acknowledge the financial support from the Ministry of Education, Science, and Technological Development of the Republic of Serbia, Grant No. 451-03-136/2025-03/200017 (“Vinca” Institute of Nuclear Sciences, University of Belgrade) and the Projects within the Cooperation Agreement between the JINR, Dubna, Russian Federation and the Republic of Serbia (P02).
\end{acknowledgments}


\appendix

\section{Supercurrent calculation}
\label{Js}

We will perform a perturbative analysis of SFS $\varphi _0$ JJ  on the main Buzdin step within the RCSJ model where the following equation describes the system dynamics: 

\begin{equation}
\beta_c \ddot{\varphi} + \dot{\varphi} + sin(\varphi -r m_y ) = I + r\dot{m}_y.
\label{CVC}
\end{equation}
In the regime $\beta_c I \gg 1 $, the harmonic solution for $\varphi$  can be written in the form:

\begin{eqnarray}
\varphi(t) &=&\phi_0 + \omega_J t +\sum_{n} \left[ a_n \cos(n \omega_J t ) + b_n \sin(n \omega_J t )\right], \nonumber \\
\dot{\varphi}(t) &=& \omega_J + \omega_J\sum_{n} n\left[ - a_n  \sin(n \omega_J t ) + b_n \cos(n \omega_J t )\right], \nonumber \\
 \ddot{\varphi}(t) &=& - \omega_J^2\sum_{n} n^2\left[ a_n  \cos(n \omega_J t ) + b_n \sin(n \omega_J t )\right],
\label{phi-harmonicA}
\end{eqnarray}
where $\phi_0 $ is initial phase and $\omega_J$ is Josephson frequency.
The current in (\ref{CVC}) can be expand as: 

\begin{eqnarray}
 I = \omega_J + \overline{j}^{(1)} + ...
\label{current-expansionA}
\end{eqnarray}
If we assume that the term $\beta_c \ddot{\varphi}$ in (\ref{CVC}) is much larger then $\dot{\varphi}$ or $sin(\varphi -r m_y )$, the last two terms will consist only of linear terms where $\varphi(t) = \phi_0 + \omega_J t$ and $\dot{\varphi}=\omega_J$.
Substituting (\ref{phi-harmonic}) into (\ref{CVC}), and taking into account only $1$-st order term ($n=1$) we obtain:

\begin{eqnarray}
-\beta_c \omega_J^2 \left[ a_1  \cos(\omega_J t ) + b_1 \sin(\omega_J t )\right] + \omega_J  = \\
= I - \sin\left(\phi_0 + \omega_J t - r m_y(t) \right) +r\dot{m}_y,\nonumber
\label{5}
\end{eqnarray}
which after substituting (\ref{current-expansion}) becomes:

\begin{eqnarray}
-\beta_c \omega_J^2 \left[ a_1  \cos(\omega_J t ) + b_1 \sin(\omega_J t )\right]  = \\
=\overline{j}^{(1)} - \sin(\phi_0 +\omega_Jt - r m_y(t) ) +r\dot{m}_y.\nonumber
\label{33}
\end{eqnarray}
Since $rm_y<<1$ we can write
\begin{eqnarray}
&-&\beta_c \omega_J^2 \left[ a_1  \cos(\omega_J t ) + b_1 \sin(\omega_J t )\right] - \overline{j}^{(1)}  =  \\
&=& - \sin(\phi_0 +\omega_Jt )+ rm_y(t)\cos(\phi_0 +\omega_Jt)  +r\dot{m}_y.\nonumber
\label{34}
\end{eqnarray}

To evaluate the size of the Buzdin step, 
we need to calculate the constant correction to the current $\overline{j}^{(1)}$.  
So first, we need to determine an average over time
of equation (\ref{34}).
Since the average from the first term on the right side $\left<- \sin(\phi_0 + \omega_Jt )\right>=0$, it remains to calculate the terms $\overline{rm_y(t)\cos(\phi_0 +\omega_Jt)}$ and $\overline{r\dot{m}_y}$.

\vspace{0.5cm}

\section{$m_y$ calculation}
\label{my}

All the calculations will be performed in the following limits:

\begin{eqnarray}
\alpha &\ll & 1, \hspace{1cm} \Rightarrow \hspace{1cm} \frac{\omega_{F}}{(1+\alpha^{2})} \approx  \omega_{F},  \nonumber \\
r &\ll & 1, \nonumber \\
m_x &\ll &1,  \nonumber \\
m_y &\ll & 1, \nonumber \\
m_z &\rightarrow & 1, \\
h_{y} & \approx &  rG \sin(\varphi ),  \nonumber \\ 
h_{z} & \approx & 1 + \dfrac{h_R}{\omega_{F}} \sin(\omega_R t). \nonumber
\label{limits}
\end{eqnarray}
After substitution of these conditions into the LLG equations (\ref{LLGJ}) one can get:

\begin{widetext}
\begin{subequations} \label{LLG_components_limit}
\begin{eqnarray}
	\dot{m_{x}} + \omega_{F} \left(1 + \dfrac{h_R}{\omega_{F}} \sin(\omega_R t)\right) m_y &=& \omega_{F} rG \sin(\omega_J t + \phi_0  ) ,\\
	\dot{m_{y}} - \omega_{F} \left(1 + \dfrac{h_R}{\omega_{F}} \sin(\omega_R t)\right) m_x  &=& 0,
\end{eqnarray}
\end{subequations}
\end{widetext}
Since for the JJ equation, we need only $m_y$-component, we can determine $\dot{m_{x}}$ from (\ref{LLG_components_limit}b), which after substitution into (\ref{LLG_components_limit}a) leads to:
\begin{eqnarray}
	\ddot{m_{y}}- & &\dfrac{\dot{m_{y}}h_R\omega_R \cos(\omega_R t)}{ \left(\omega_{F}  + h_R \sin(\omega_R t)\right)} +  \left(\omega_{F} + h_R \sin(\omega_R t)\right)^2 m_y = \nonumber  \\
& & =\omega_{F} rG \sin(\omega_J t + \phi_0   )\left(\omega_{F} + h_R \sin(\omega_R t)\right).
 \label{m_y_dyn}
\end{eqnarray}
The homogeneous solution of this equation is given by:
\begin{eqnarray} \label{homog_sol2}
	  m_y &=&C_1 \sin\left(\omega_{F}t  - \dfrac{h_R}{\omega_{R}} \cos(\omega_R t)\right) + \\
   && C_2 \cos\left(\omega_{F}t  - \dfrac{h_R}{\omega_{R}} \cos(\omega_R t)\right), \nonumber   
\end{eqnarray}
meanwhile, the inhomogeneous one can be found by applying a method for varying arbitrary constant in the following form:
\begin{eqnarray} \label{inhomog_sol}
	  m_y &=& \int_{t_0}^{t} \dfrac{y_1(\xi)y_2(t)-y_2(\xi)y_1(t)}{W(\xi)}R(\xi)d\xi,   
\end{eqnarray}
where 
\begin{eqnarray}\label{y_sol}
	  y_1(t) &=& \sin\left(\omega_{F}t  - \dfrac{h_R}{\omega_{R}} \cos(\omega_R t)\right), \nonumber\\
	  y_2(t) &=& \cos\left(\omega_{F}t  - \dfrac{h_R}{\omega_{R}} \cos(\omega_R t)\right), \\
	  W(\xi) &=& y_1(\xi)y'_2(\xi)-y'_1(\xi)y_2(\xi)= \nonumber\\ 
            &=&-\omega_{F}-h_R \sin(\omega_R \xi) ,  \nonumber\\
	  R(\xi) &=& \omega_{F} rG \sin(\omega_J t + \phi_0  )\left(\omega_{F} + h_R \sin(\omega_R \xi)\right). \nonumber
\end{eqnarray}
Substituting (\ref{y_sol}) into (\ref{inhomog_sol}) and taking into account that $\sin\alpha=Re(-ie^{i\alpha})$ and $\cos\alpha=Re(e^{i\alpha})$ one can get:
\begin{widetext}
\begin{eqnarray}
	  m_y &=& \dfrac{1}{2}\omega_{F} rG Re\left(\sin(\omega_{F}t  - \dfrac{h_R}{\omega_{R}} \cos(\omega_R t)) \int_{t_0}^{t}d\xi \left[(-i)e^{i(\omega_J \xi + \phi_0  -\omega_{F}\xi  + \frac{h_R}{\omega_{R}} \cos(\omega_R \xi))}+  
	  ie^{i(-\omega_J \xi - \phi_0 -\omega_{F}\xi + \frac{h_R}{\omega_{R}} \cos(\omega_R \xi))}\right] - \right. \nonumber \\ 
	  &-&\left. \cos(\omega_{F}t  - \dfrac{h_R}{\omega_{R}} \cos(\omega_R t)) \int_{t_0}^{t}d\xi \left[e^{i(\omega_J \xi + \phi_0 -\omega_{F}\xi  + \frac{h_R}{\omega_{R}} \cos(\omega_R \xi))}-  
	  -e^{i(-\omega_J \xi - \phi_0 -\omega_{F}\xi  + \frac{h_R}{\omega_{R}} \cos(\omega_R \xi))}\right]\right).   
\end{eqnarray}
\end{widetext}
Then using $e^{iz \cos\theta} = \sum^{\infty}_{n=-\infty}J_n(z)e^{in\theta}$ we obtain:
\begin{widetext}
\begin{eqnarray} \label{inhomog_sol_my1}
	  m_y &=& \dfrac{1}{2}\omega_{F} rG Re\left(e^{i(\omega_{F}t  - \frac{h_R}{\omega_{R}} \cos(\omega_R t))} \int_{t_0}^{t}d\xi \sum^{\infty}_{n=-\infty}i^nJ_n\left(\dfrac{h_R}{\omega_{R}}\right)\left[ e^{i( n\omega_R \xi-\omega_{F}\xi -\omega_J \xi - \phi_0 )}-  
	  e^{i(n\omega_R \xi-\omega_{F}\xi + \omega_J \xi + \phi_0 )}\right] \right). 
\end{eqnarray}
\end{widetext}

\begin{figure}[tbh]
	\centering
	\includegraphics[width=0.6\linewidth]{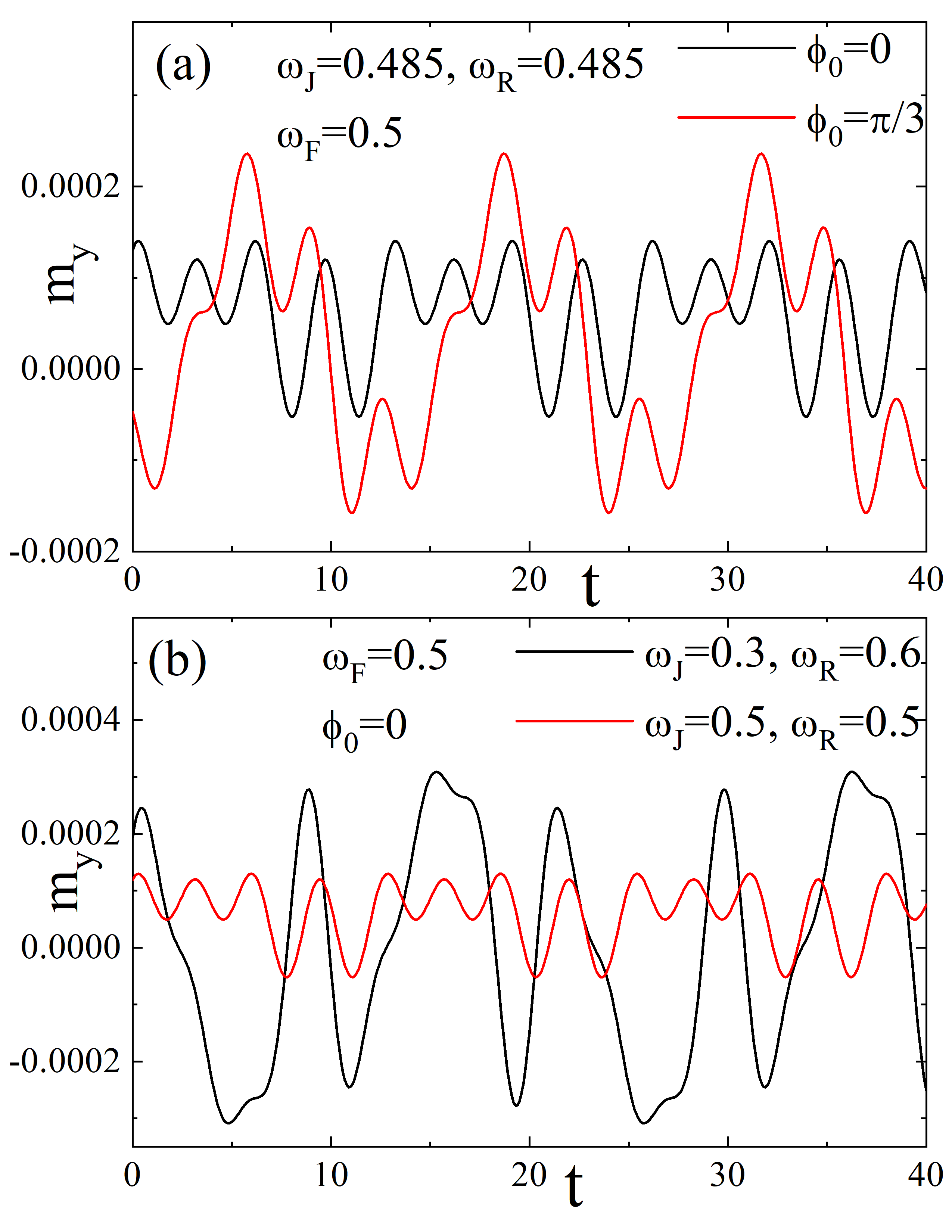}
	\caption{(a) Oscillations of the magnetic component $m_y$ at the Buzdin step $\omega _{J}=0.485$ for $\omega _R=0.485$, $\omega _F=0.5$ and two different values of $\phi_0 =0, \pi /3$.
    (b) Oscillations of the magnetic component $m_y$ at the Buzdin step $\omega _{J}=0.3$ for $\omega _R=0.6$, and at $\omega _{J}=0.5$ for $\omega _R=0.5$ where $\omega _F=0.5$ and $\phi_0 =0$.
    The rest of the parameters is the same as in Fig. \ref{Fig2}.   
    }
	\label{Fig8}
\end{figure}

After integration, one can get:
\begin{widetext}
\begin{eqnarray}
	  m_y &=& \dfrac{rG\omega_{F} }{2} Re\left(  \sum^{\infty}_{n=-\infty}i^{n-1}J_n\left(\dfrac{h_R}{\omega_{R}}\right)\left[\dfrac{e^{i( n\omega_R t-\frac{h_R}{\omega_{R}} \cos(\omega_R t) - \omega_J t-\phi_0 )}}{ n\omega_R -\omega_{F} - \omega_J } - \dfrac{e^{i(n\omega_R t-\frac{h_R}{\omega_{R}} \cos(\omega_R t) +  \omega_J t+\phi_0 )}}{n\omega_R -\omega_{F} +  \omega_J} \nonumber 
	  + C_ne^{i(\omega_{F}t  - \frac{h_R}{\omega_{R}} \cos(\omega_R t))}\right] \right),\nonumber \\
     & &
\end{eqnarray}
\end{widetext}
where by substituting the initial condition $m_y=0$ one can calculate that $C_n=0$ and after using the expansion $e^{-i\frac{h_R}{\omega_{R}} \cos(\omega_R t)} = \sum^{\infty}_{k=-\infty}i^kJ_k(-\frac{h_R}{\omega_{R}})e^{ik\omega_R t}$, taking into account that $J_k(-\delta)=(-1)^kJ_k(\delta)$ and introducing $q=~n+k$, $m_y$ is given by: 
\begin{widetext}
\begin{eqnarray}\label{phy_m_int}
	  m_y &=& \dfrac{rG\omega_{F} }{2} Re\left(  \sum^{\infty}_{n=-\infty}i^{n-1}J_n(\dfrac{h_R}{\omega_{R}})\left[\sum^{\infty}_{q=-\infty}i^{q-n}(-1)^{q-n}J_{q-n}\left(\frac{h_R}{\omega_{R}}\right)\dfrac{e^{i( q\omega_R t - \omega_J t-\phi_0 )}}{ n\omega_R -\omega_{F} - \omega_J } - \right. \right.\\ \nonumber
	 & & \left. \left. \sum^{\infty}_{q=-\infty}i^{q-n}(-1)^{q-n}J_{q-n}\left(\frac{h_R}{\omega_{R}}\right)\dfrac{e^{i(q\omega_R t +  \omega_J t+\phi_0 )}}{n\omega_R -\omega_{F} +  \omega_J}\right] \right).\nonumber
\end{eqnarray}
\end{widetext}
In Fig. \ref{Fig8} we present the examples of $m_y(t)$ dependence on Buzdin steps for different values of the initial phase ($\phi_0 $) in (a) and for different values of driving frequencies $\omega_R$ (b). 
As Fig. \ref{Fig8} (a) demonstrates, the nonzero Josephson phase $\ phi_0$ does not affect the period but increases the amplitude of the oscillations.
On the other hand, in Fig. \ref{Fig8} (b), changing the driving frequency $\omega _R$ determines the position of Buzdin steps and changes the amplitude and the period of the    $m_y(t)$ oscillations.

\section{$\overline{rm_y(t)\cos(\omega_Jt+\phi_0 )}$ calculation}
\label{rmycos}

Let us first calculate the term $\overline{rm_y(t)\cos(\omega_Jt+\phi_0 )}$.
Taking into account that $\cos(\omega_Jt+\phi_0 )=\dfrac{e^{i\omega_Jt+i\phi_0 }+e^{-i\omega_Jt-i\phi_0 }}{2}$ we can write:
\begin{widetext}
\begin{eqnarray}\label{phimcosC}
	  rm_y(t)\cos(\omega_Jt+\phi_0 ) &=& \dfrac{r^2G\omega_{F} }{4} Re\left(  \sum^{\infty}_{n=-\infty}i^{n-1}J_n\left(\dfrac{h_R}{\omega_{R}}\right)\left[\sum^{\infty}_{q=-\infty}i^{q-n}(-1)^{q-n}J_{q-n}\left(\frac{h_R}{\omega_{R}}\right)\dfrac{e^{i( q\omega_R t )}}{ n\omega_R -\omega_{F} - \omega_J } -\right. \right.\nonumber \\ 
	  &-& \sum^{\infty}_{q=-\infty}i^{q-n}(-1)^{q-n}J_{q-n}\left(\frac{h_R}{\omega_{R}}\right)\dfrac{e^{i(q\omega_R t )}}{n\omega_R -\omega_{F} +  \omega_J} +
	  \sum^{\infty}_{q=-\infty}i^{q-n}(-1)^{q-n}J_{q-n}\left(\frac{h_R}{\omega_{R}}\right)\dfrac{e^{i( q\omega_R t - 2\omega_J t-2\phi_0 )}}{ n\omega_R -\omega_{F} - \omega_J } - \nonumber \\ 
	  &-& \left. \left. \sum^{\infty}_{q=-\infty}i^{q-n}(-1)^{q-n}J_{q-n}\left(\frac{h_R}{\omega_{R}}\right)\dfrac{e^{i(q\omega_R t + 2 \omega_J t+2\phi_0 )}}{n\omega_R -\omega_{F} +  \omega_J} \right]\right).\nonumber
\end{eqnarray}
\end{widetext}
The first two terms in Eq. (\ref{phimcosC}) do not have the average real parts, since to have the time independence in the exponents we need to imply 
$ q\omega_R t=0 \Rightarrow q=0$ 
and then 
$i^{n-1}i^{q-n}=i^{n-1}i^{-n}=i^{-1}$, so the $Re(i^{-1}x)=0$. 
Therefore, the only terms that remain are the fourth and the fifth ones:
\begin{widetext}
\begin{eqnarray}\label{cos_termD}
	  rm_y(t)\cos(\omega_Jt+\phi_0 ) &=& \dfrac{r^2G\omega_{F} }{4} Re\left(  \sum^{\infty}_{n=-\infty}i^{n-1}J_n\left(\dfrac{h_R}{\omega_{R}}\right)\left[ \sum^{\infty}_{q=-\infty}i^{q-n}(-1)^{q-n}J_{q-n}\left(\frac{h_R}{\omega_{R}}\right)\dfrac{e^{i( q\omega_R t - 2\omega_J t-2\phi_0 )}}{ n\omega_R -\omega_{F} - \omega_J } -\right. \right.\nonumber \\ 
	  &-& \left. \left. \sum^{\infty}_{q=-\infty}i^{q-n}(-1)^{q-n}J_{q-n}\left(\frac{h_R}{\omega_{R}}\right)\dfrac{e^{i(q\omega_R t + 2 \omega_J t+2\phi_0 )}}{n\omega_R -\omega_{F} +  \omega_J} \right]\right).
\end{eqnarray}
\end{widetext}

\section{$\overline{r\dot{m}_y}$ calculation}
\label{rmydot}

Now we can calculate $r\dot{m}_y$, from (\ref{phy_m_int}) we can write:
\begin{widetext}
\begin{eqnarray}\label{phy_dot}
	 r \dot{m}_y &=&  \dfrac{r^2G\omega_{F} }{2} Re\left(  \sum^{\infty}_{n=-\infty}i^{n-1}J_n\left(\dfrac{h_R}{\omega_{R}}\right) \times  
	   \left[\sum^{\infty}_{q=-\infty}i^{q-n+1}(-1)^{q-n}J_k\left(\frac{h_R}{\omega_{R}}\right)\dfrac{ q\omega_R - \omega_J}{ n\omega_R -\omega_{F} - \omega_J }e^{i( q\omega_R t - \omega_J t-\phi_0 )} - \right. \right. \nonumber \\ 
	  &-& \left. \left. \sum^{\infty}_{q=-\infty}i^{q-n+1}(-1)^{q-n}J_{q-n}\left(\frac{h_R}{\omega_{R}}\right)\dfrac{q\omega_R +  \omega_J}{n\omega_R -\omega_{F} +  \omega_J}e^{i(q\omega_R t +  \omega_J t+\phi_0 )} \right] \right).
\end{eqnarray}
\end{widetext}
For the time-independent term to appear the condition $q - 1=0$ or $q + 1=0$ must be satisfied.
However, both of those conditions bring a zero in the numerator. Therefore, we must have $\overline{r\dot{m}_y}=0$.

The only nonzero term will come from Eq.(\ref{cos_termD}): 
\begin{widetext}
\begin{eqnarray}\label{cos_term_F}
	  rm_y(t)\cos(\omega_Jt+\phi_0 ) &=& \dfrac{r^2G\omega_{F} }{4} Re\left(  \sum^{\infty}_{n=-\infty}i^{n-1}J_n\left(\dfrac{h_R}{\omega_{R}}\right)\left[ \sum^{\infty}_{q=-\infty}i^{q-n}(-1)^{q-n}J_{q-n}\left(\frac{h_R}{\omega_{R}}\right)\dfrac{e^{i( (q - 2)\omega_R t-2\phi_0 )}}{ (n-1)\omega_R -\omega_{F} } - \right. \right. \nonumber \\
	  & - & \left. \left. \sum^{\infty}_{q=-\infty}i^{q-n}(-1)^{q-n}J_{q-n}\left(\frac{h_R}{\omega_{R}}\right)\dfrac{e^{i((q + 2)\omega_R t+2\phi_0 )}}{(n+1)\omega_R -\omega_{F}} \right]\right).
\end{eqnarray}
\end{widetext}
Introducing $q=n+k$ so that $k=q-n$, 
Eq.(\ref{cos_term_F}) becomes:

\begin{widetext}
\begin{eqnarray}\label{cos_termFq}
	  rm_y(t)\cos(\omega_Jt+\phi_0 ) &=& \dfrac{r^2G\omega_{F} }{4} Re\left(  \sum^{\infty}_{n=-\infty}i^{n-1}J_n\left(\dfrac{h_R}{\omega_{R}}\right)\left[ \sum^{\infty}_{q=-\infty}i^{q-n}(-1)^{q-n}J_{q-n}\left(\frac{h_R}{\omega_{R}}\right)\dfrac{e^{i( (q - 2)\omega_R t-2\phi_0 )}}{ (n-1)\omega_R -\omega_{F} } - \right. \right. \nonumber \\
	  & - & \left. \left. \sum^{\infty}_{k=-\infty}i^{q-n}(-1)^{q-n}J_{q-n}\left(\frac{h_R}{\omega_{R}}\right)\dfrac{e^{i((q + 2)\omega_R t+2\phi_0 )}}{(n+1)\omega_R -\omega_{F}} \right]\right).
\end{eqnarray}
\end{widetext}

\section{Zero voltage state}
\label{ZVS}

There is no Josephson oscillation in the zero voltage state. So, $\omega_J = 0$ and Eq.(\ref{cos_termD}) is given by:
\begin{widetext}
\begin{eqnarray}\label{cos_term_Ic}
	 rm_y(t)\cos(\phi_0 ) =& & \dfrac{r^2G\omega_{F} }{4} Re\left(  \sum^{\infty}_{n=-\infty}i^{n-1}J_n\left(\dfrac{h_R}{\omega_{R}}\right) \left[ \sum^{\infty}_{q=-\infty}i^{q-n}(-1)^{q-n}J_{q-n}\left(\frac{h_R}{\omega_{R}}\right) \dfrac{e^{i( q\omega_R t-2\phi_0  )}}{ n\omega_R -\omega_{F}  } - \right. \right. \\ 
	& - & \left. \left. \sum^{\infty}_{q=-\infty}i^{q-n}(-1)^{q-n}J_{q-n}\left(\frac{h_R}{\omega_{R}}\right)\dfrac{e^{i(q\omega_R t+2\phi_0  )}}{n\omega_R -\omega_{F} } \right]\right) . \nonumber
\end{eqnarray}
\end{widetext}

These two terms do not have the average real parts, since to have the time independence in the exponents we need to imply 
$q\omega_R t=0 \Rightarrow q=0$ 
and then 
$i^{n-1}i^{q-n}=i^{n-1}i^{-n}=i^{-1}$, so the $Re(i^{-1}x)=0$.  
Therefore, there is no contribution to the zero voltage state from the magnetic subsystem.

\section{The Buzdin step for $\omega_J=\omega_R \sim \omega_F$}
\label{BSwF}

Taking into account that we need only the largest terms in Eq. (\ref{cos_termFq}), which will be near the resonance condition $\omega_R -\omega_{F} \ll 1$, and that the denominator should be equal to $\omega_R -\omega_{F}$, we obtain the following conditions for $q$ and $n$:

\begin{eqnarray}
	  q - 2 &=& 0, \hspace{1cm} n - 1 = 1, \\
	  q + 2 &=& 0, \hspace{1cm} n + 1 = 1.
        \nonumber
\end{eqnarray}
Then the pairs of $q$ and $n$ for the first and second terms are the following:

\begin{eqnarray}
	  n &=& 2, \hspace{1cm} q = 2, \\
	  n &=& 0, \hspace{1cm} q = -2. \nonumber
\end{eqnarray}
Substituting those values into (\ref{cos_termFq}) and taking into account that $\dfrac{e^{i2\phi_0 }+e^{ -i2\phi_0 }}{2i}=\sin(2\phi_0 )$ one can get:

\begin{multline}
\label{cos_term_4}
\overline{ rm_y(t)\cos(\omega_Jt+\phi_0 )} = \\ 
     =\dfrac{r^2G\omega_{F} }{2(\omega_R -\omega_{F})} J_2\left(\dfrac{h_R}{\omega_{R}}\right)J_0\left(\dfrac{h_R}{\omega_{R}}\right)\sin(2\phi_0 ).
\end{multline}

From the above results, the average supercurrent can be determined as
\begin{widetext}
\begin{eqnarray}\label{barj}
\overline{j}^{(1)}  =   \overline{rm_y(t)\cos(\omega_Jt+\phi_0 )}  =\dfrac{r^2G\omega_{F} }{2(\omega_R -\omega_{F})} J_2\left(\dfrac{h_R}{\omega_{R}}\right)J_0\left(\dfrac{h_R}{\omega_{R}}\right)\sin(2\phi_0 ),
\label{add_current}
\end{eqnarray}
\end{widetext}
and from (\ref{barj}) the step width calculated as:
\begin{widetext}
\begin{eqnarray}
\Delta j^{(1)}  = max\left(\overline{j}^{(1)}\right)-min\left(\overline{j}^{(1)}\right) =  \dfrac{r^2G\omega_{F} }{(\omega_R -\omega_{F})} J_2\left(\dfrac{h_R}{\omega_{R}}\right)J_0\left(\dfrac{h_R}{\omega_{R}}\right).
\label{step_width}
\end{eqnarray}
\end{widetext}

\section{The Buzdin step for $\omega_J=\omega_R \sim \frac{\omega_F}{2}$}
\label{BSwf/2}

In this case, in Eq. (\ref{cos_termFq}), we consider only the terms, that are near the resonance condition $\omega_R -\dfrac{\omega_{F}}{2}\ll 1$.  
This leads to the following conditions on $q$ and $n$:

\begin{subequations} 
\begin{eqnarray}
	  q - 2 &=& 0, \hspace{1cm} n-1 = 2 , \nonumber \\
	  q + 2 &=& 0, \hspace{1cm} n+1 = 2, \nonumber
\end{eqnarray}
\end{subequations}

So, the pairs of $q$ and $n$ for the first and second term are following:

\begin{subequations} 
\begin{eqnarray}
	  n &=& 3, \hspace{1cm} q = 2, \nonumber \\
	  n &=& 1, \hspace{1cm} q = -2, \nonumber
\end{eqnarray}
\end{subequations}

Following the same procedure as in Sec. \ref{BSwF} we obtain:
\begin{multline}
\label{cos_term_wf/2}
\overline{ rm_y(t)\cos(\omega_Jt+\phi_0 )} = \\ 
     =- \dfrac{r^2G\omega_{F} }{2(2\omega_R -\omega_{F})} J_3\left(\dfrac{h_R}{\omega_{R}}\right)J_1\left(\dfrac{h_R}{\omega_{R}}\right)\sin(2\phi_0 ),
\end{multline}
which gives us the average supercurrent:
\begin{widetext}
\begin{eqnarray}\label{barjwf/2}
\overline{j}^{(1)}  =-\dfrac{r^2G\omega_{F} }{2(2\omega_R -\omega_{F})} J_3\left(\dfrac{h_R}{\omega_{R}}\right)J_1\left(\dfrac{h_R}{\omega_{R}}\right)\sin(2\phi_0 ),
\label{add_current}
\end{eqnarray}
\end{widetext}
and from there, the BS width:
\begin{eqnarray}
\Delta j^{(1)}  =  \dfrac{r^2G\omega_{F} }{2\omega_R -\omega_{F}} J_3\left(\dfrac{h_R}{\omega_{R}}\right)J_1\left(\dfrac{h_R}{\omega_{R}}\right).
\label{step_widthwf/2}
\end{eqnarray}

\bibliography{SFSJJh}{}

\begin{thebibliography}{42}
\providecommand{\natexlab}[1]{#1}
\providecommand{\url}[1]{\texttt{#1}}
\expandafter\ifx\csname urlstyle\endcsname\relax
  \providecommand{\doi}[1]{doi: #1}\else
  \providecommand{\doi}{doi: \begingroup \urlstyle{rm}\Url}\fi

\bibitem[Linder and Robinson(2015)]{linder2015superconducting}
J.~Linder and J.~W.~A. Robinson.
\newblock Superconducting spintronics.
\newblock \emph{Nature Physics}, 11\penalty0 (4):\penalty0 307--315, 2015.

\bibitem[Eschrig(2011)]{eschrig2011spin}
M.~Eschrig.
\newblock Spin-polarized supercurrents for spintronics.
\newblock \emph{Physics Today}, 64\penalty0 (1):\penalty0 43--49, 2011.

\bibitem[Golubov and Kupriyanov(2017)]{golubov2017controlling}
Alexander~A Golubov and Mikhail~Yu Kupriyanov.
\newblock Controlling magnetism.
\newblock \emph{Nature materials}, 16\penalty0 (2):\penalty0 156--157, 2017.

\bibitem[Mel’nikov et~al.(2022)Mel’nikov, Mironov, Samokhvalov, and
  Buzdin]{mel2022superconducting}
AS~Mel’nikov, Sergei~Viktorovich Mironov, Aleksei~Vladimirovich Samokhvalov,
  and Alexander~Ivanovich Buzdin.
\newblock Superconducting spintronics: state of the art and prospects.
\newblock \emph{Uspekhi Fiz. Nauk}, 192:\penalty0 1339--1384, 2022.

\bibitem[Ryazanov et~al.(2012)Ryazanov, Bol’ginov, Sobanin, Vernik, Tolpygo,
  Kadin, and Mukhanov]{ryazanov2012magnetic}
Valery~V Ryazanov, Vitaly~V Bol’ginov, Danila~S Sobanin, Igor~V Vernik,
  Sergey~K Tolpygo, Alan~M Kadin, and Oleg~A Mukhanov.
\newblock Magnetic josephson junction technology for digital and memory
  applications.
\newblock \emph{Physics Procedia}, 36:\penalty0 35--41, 2012.

\bibitem[Cai et~al.(2023)Cai, {\v{Z}}uti{\'c}, and Han]{cai2023superconductor}
R.~Cai, I.~{\v{Z}}uti{\'c}, and W.~Han.
\newblock Superconductor/ferromagnet heterostructures: a platform for
  superconducting spintronics and quantum computation.
\newblock \emph{Advanced Quantum Technologies}, 6\penalty0 (1):\penalty0
  2200080, 2023.

\bibitem[Guarcello and Bergeret(2020)]{guarcello2020cryogenic}
Claudio Guarcello and FS~Bergeret.
\newblock Cryogenic memory element based on an anomalous josephson junction.
\newblock \emph{Physical Review Applied}, 13\penalty0 (3):\penalty0 034012,
  2020.

\bibitem[Konschelle and Buzdin(2009)]{konschelle2009magnetic}
F.~Konschelle and A.~I. Buzdin.
\newblock Magnetic moment manipulation by a josephson current.
\newblock \emph{Physical Review Letters}, 102\penalty0 (1):\penalty0 017001,
  2009.

\bibitem[Buzdin(2008)]{buzdin2008direct}
A.~I. Buzdin.
\newblock Direct coupling between magnetism and superconducting current in the
  josephson $\varphi$ 0 junction.
\newblock \emph{Physical review letters}, 101\penalty0 (10):\penalty0 107005,
  2008.

\bibitem[Buzdin(2005)]{buzdin2005proximity}
A.~I. Buzdin.
\newblock Proximity effects in superconductor-ferromagnet heterostructures.
\newblock \emph{Reviews of modern physics}, 77\penalty0 (3):\penalty0 935,
  2005.

\bibitem[Shukrinov(2022)]{shukrinov2022anomalous}
Yu.~M. Shukrinov.
\newblock Anomalous josephson effect.
\newblock \emph{Physics-Uspekhi}, 65\penalty0 (4):\penalty0 317, 2022.

\bibitem[Amundsen et~al.(2024)Amundsen, Linder, Robinson, {\v{Z}}uti{\'c}, and
  Banerjee]{amundsen2024colloquium}
Morten Amundsen, Jacob Linder, Jason~WA Robinson, Igor {\v{Z}}uti{\'c}, and
  Niladri Banerjee.
\newblock Colloquium: Spin-orbit effects in superconducting hybrid structures.
\newblock \emph{Reviews of Modern Physics}, 96\penalty0 (2):\penalty0 021003,
  2024.

\bibitem[Szombati et~al.(2016)Szombati, Nadj-Perge, Car, Plissard, Bakkers, and
  Kouwenhoven]{szombati2016josephson}
DB~Szombati, S~Nadj-Perge, Diane Car, SR~Plissard, EPAM Bakkers, and
  LP~Kouwenhoven.
\newblock Josephson $\phi$ 0-junction in nanowire quantum dots.
\newblock \emph{Nature Physics}, 12\penalty0 (6):\penalty0 568--572, 2016.

\bibitem[Assouline et~al.(2019)Assouline, Feuillet-Palma, Bergeal, Zhang,
  Mottaghizadeh, Zimmers, Lhuillier, Eddrie, Atkinson, Aprili,
  et~al.]{assouline2019spin}
A.~Assouline, C.~Feuillet-Palma, N.~Bergeal, T.~Zhang, A.~Mottaghizadeh,
  A.~Zimmers, E.~Lhuillier, M.~Eddrie, P.~Atkinson, M.~Aprili, et~al.
\newblock Spin-orbit induced phase-shift in bi2se3 josephson junctions.
\newblock \emph{Nature communications}, 10\penalty0 (1):\penalty0 126, 2019.

\bibitem[Mayer et~al.(2020)Mayer, Dartiailh, Yuan, Wickramasinghe, Rossi, and
  Shabani]{mayer2020gate}
William Mayer, Matthieu~C Dartiailh, Joseph Yuan, Kaushini~S Wickramasinghe,
  Enrico Rossi, and Javad Shabani.
\newblock Gate controlled anomalous phase shift in al/inas josephson junctions.
\newblock \emph{Nature communications}, 11\penalty0 (1):\penalty0 212, 2020.

\bibitem[Takahashi et~al.(2007)Takahashi, Hikino, Mori, Martinek, and
  Maekawa]{takahashi2007supercurrent}
S~Takahashi, S~Hikino, M~Mori, J~Martinek, and S~Maekawa.
\newblock Supercurrent pumping in josephson junctions with a half-metallic
  ferromagnet.
\newblock \emph{Physical review letters}, 99\penalty0 (5):\penalty0 057003,
  2007.

\bibitem[Rabinovich et~al.(2019)Rabinovich, Bobkova, Bobkov, and
  Silaev]{rabinovich2019resistive}
D.~S. Rabinovich, I.~V. Bobkova, A.~M. Bobkov, and M.~A. Silaev.
\newblock Resistive state of superconductor-ferromagnet-superconductor
  josephson junctions in the presence of moving domain walls.
\newblock \emph{Physical Review Letters}, 123\penalty0 (20):\penalty0 207001,
  2019.

\bibitem[Pal and Benjamin(2018)]{pal2018quantized}
Subhajit Pal and Colin Benjamin.
\newblock Quantized josephson phase battery.
\newblock \emph{arXiv preprint arXiv:1811.02484}, 2018.

\bibitem[Trahms et~al.(2023)Trahms, Melischek, Steiner, Mahendru, Tamir,
  Bogdanoff, Peters, Reecht, Winkelmann, von Oppen, et~al.]{trahms2023diode}
Martina Trahms, Larissa Melischek, Jacob~F Steiner, Bharti Mahendru, Idan
  Tamir, Nils Bogdanoff, Olof Peters, Ga{\"e}l Reecht, Clemens~B Winkelmann,
  Felix von Oppen, et~al.
\newblock Diode effect in josephson junctions with a single magnetic atom.
\newblock \emph{Nature}, 615\penalty0 (7953):\penalty0 628--633, 2023.

\bibitem[Narita et~al.(2022)Narita, Ishizuka, Kawarazaki, Kan, Shiota,
  Moriyama, Shimakawa, Ognev, Samardak, Yanase, et~al.]{narita2022field}
Hideki Narita, Jun Ishizuka, Ryo Kawarazaki, Daisuke Kan, Yoichi Shiota,
  Takahiro Moriyama, Yuichi Shimakawa, Alexey~V Ognev, Alexander~S Samardak,
  Youichi Yanase, et~al.
\newblock Field-free superconducting diode effect in noncentrosymmetric
  superconductor/ferromagnet multilayers.
\newblock \emph{Nature Nanotechnology}, 17\penalty0 (8):\penalty0 823--828,
  2022.

\bibitem[Botha et~al.(2023)Botha, Shukrinov, Teki{\'c}, and
  Kolahchi]{botha2023chaotic}
AE~Botha, Yu~M Shukrinov, J~Teki{\'c}, and MR~Kolahchi.
\newblock Chaotic dynamics from coupled magnetic monodomain and josephson
  current.
\newblock \emph{Physical Review E}, 107\penalty0 (2):\penalty0 024205, 2023.

\bibitem[Abdelmoneim et~al.(2022)Abdelmoneim, Shukrinov, Kulikov, ElSamman, and
  Nashaat]{abdelmoneim2022locking}
S.~A. Abdelmoneim, Yu.~M. Shukrinov, K.~V. Kulikov, H.~ElSamman, and
  M.~Nashaat.
\newblock Locking of magnetization and josephson oscillations at ferromagnetic
  resonance in a $\varphi$ 0 junction under external radiation.
\newblock \emph{Physical Review B}, 106\penalty0 (1):\penalty0 014505, 2022.

\bibitem[Shukrinov et~al.(2021)Shukrinov, Rahmonov, Janalizadeh, and
  Kolahchi]{shukrinov2021anomalous}
Yu.~M. Shukrinov, I.~R. Rahmonov, A.~Janalizadeh, and M.~R. Kolahchi.
\newblock Anomalous gilbert damping and duffing features of the
  superconductor-ferromagnet-superconductor $\varphi$ 0 josephson junction.
\newblock \emph{Physical Review B}, 104\penalty0 (22):\penalty0 224511, 2021.

\bibitem[Shukrinov et~al.(2024)Shukrinov, Kovalenko, Teki{\'c}, Kulikov, and
  Nashaat]{shukrinov2024buzdin}
Yu~M Shukrinov, E~Kovalenko, Jasmina Teki{\'c}, K~Kulikov, and M~Nashaat.
\newblock Buzdin, shapiro, and chimera steps in $\varphi$ 0 josephson
  junctions.
\newblock \emph{Physical Review B}, 109\penalty0 (2):\penalty0 024511, 2024.

\bibitem[Bobkova et~al.(2022)Bobkova, Bobkov, and
  Silaev]{bobkova2022magnetoelectric}
IV~Bobkova, AM~Bobkov, and MA~Silaev.
\newblock Magnetoelectric effects in josephson junctions.
\newblock \emph{Journal of Physics: Condensed Matter}, 34\penalty0
  (35):\penalty0 353001, 2022.

\bibitem[Shapiro(1963)]{shapiro1963josephson}
Sidney Shapiro.
\newblock Josephson currents in superconducting tunneling: The effect of
  microwaves and other observations.
\newblock \emph{Physical Review Letters}, 11\penalty0 (2):\penalty0 80, 1963.

\bibitem[Kulikov et~al.(2024)Kulikov, Anghel, Nashaat, Dolineanu, Sameh, and
  Shukrinov]{kulikov2024resonance}
KV~Kulikov, DV~Anghel, M~Nashaat, M~Dolineanu, M~Sameh, and Yu~M Shukrinov.
\newblock Resonance phenomena in a nanomagnet coupled to a josephson junction
  under external periodic drive.
\newblock \emph{Physical Review B}, 109\penalty0 (1):\penalty0 014429, 2024.

\bibitem[Benz et~al.(1990)Benz, Rzchowski, Tinkham, and
  Lobb]{benz1990fractional}
SP~Benz, MS~Rzchowski, M~Tinkham, and CJ~Lobb.
\newblock Fractional giant shapiro steps and spatially correlated phase motion
  in 2d josephson arrays.
\newblock \emph{Physical review letters}, 64\penalty0 (6):\penalty0 693, 1990.

\bibitem[Sellier et~al.(2004)Sellier, Baraduc, Lefloch, and
  Calemczuk]{sellier2004half}
Hermann Sellier, Claire Baraduc, Fran{\c{c}}ois Lefloch, and Roberto Calemczuk.
\newblock Half-integer shapiro steps at the 0-$\pi$ crossover of a
  ferromagnetic josephson junction.
\newblock \emph{Physical review letters}, 92\penalty0 (25):\penalty0 257005,
  2004.

\bibitem[Panghotra et~al.(2020)Panghotra, Raes, de~Souza~Silva, Cools, Keijers,
  Scheerder, Moshchalkov, and Van~de Vondel]{panghotra2020giant}
R~Panghotra, B~Raes, Cl{\'e}cio~C de~Souza~Silva, I~Cools, W~Keijers,
  JE~Scheerder, VV~Moshchalkov, and J~Van~de Vondel.
\newblock Giant fractional shapiro steps in anisotropic josephson junction
  arrays.
\newblock \emph{Communications Physics}, 3\penalty0 (1):\penalty0 53, 2020.

\bibitem[Teki{\'c} and Mali(2016)]{tekic2016ac}
Jasmina Teki{\'c} and Petar Mali.
\newblock \emph{The ac driven Frenkel-Kontorova model}.
\newblock Institut za nuklearne nauke VIN{\v{C}}A, 2016.

\bibitem[Kammermeier and Scheer(2024)]{kammermeier2024magnetization}
Lukas Kammermeier and Elke Scheer.
\newblock Magnetization control of the critical current in a s-(s/f)-s
  superconducting switch.
\newblock \emph{Applied Physics Letters}, 124\penalty0 (16), 2024.

\bibitem[Satariano et~al.(2024)Satariano, Volkov, Ahmad, Di~Palma, Ferraiuolo,
  Vettoliere, Granata, Montemurro, Parlato, Pepe,
  et~al.]{satariano2024nanoscale}
Roberta Satariano, Anatoly~Fjodorovich Volkov, Halima~Giovanna Ahmad, Luigi
  Di~Palma, Raffaella Ferraiuolo, Antonio Vettoliere, Carmine Granata, Domenico
  Montemurro, Loredana Parlato, Giovanni~Piero Pepe, et~al.
\newblock Nanoscale spin ordering and spin screening effects in tunnel
  ferromagnetic josephson junctions.
\newblock \emph{Communications Materials}, 5\penalty0 (1):\penalty0 67, 2024.

\bibitem[Birge and Satchell(2024)]{birge2024ferromagnetic}
Norman~O Birge and Nathan Satchell.
\newblock Ferromagnetic materials for josephson $\{$$\backslash$pi$\}$
  junctions.
\newblock \emph{arXiv preprint arXiv:2401.04219}, 2024.

\bibitem[Lifshitz and Pitaevskii(1991)]{lif91course}
E.~M. Lifshitz and L.~P. Pitaevskii.
\newblock Course of theoretical physics, theory of the condensed state, 1991.

\bibitem[Tinkham(2004)]{tinkham2004introduction}
M.~Tinkham.
\newblock \emph{Introduction to superconductivity}.
\newblock Courier Corporation, 2004.

\bibitem[Shukrinov et~al.(2007)Shukrinov, Mahfouzi, and
  Pedersen]{shukrinov2007investigation}
Yu.~M. Shukrinov, F.~Mahfouzi, and N.~F. Pedersen.
\newblock Investigation of the breakpoint region in stacks with a finite number
  of intrinsic josephson junctions.
\newblock \emph{Physical Review B}, 75\penalty0 (10):\penalty0 104508, 2007.

\bibitem[Buckel and Kleiner(2008)]{buckel2008superconductivity}
W.~Buckel and R.~Kleiner.
\newblock \emph{Superconductivity: fundamentals and applications}.
\newblock John Wiley \& Sons, 2008.

\bibitem[Ya. V.~Fominov(2022)]{fominov2022}
D.~S.~Mikhailov Ya. V.~Fominov.
\newblock Asymmetric higher-harmonic squid as a josephson diode.
\newblock \emph{Physical Review B}, 106:\penalty0 134514, 2022.

\bibitem[G.~S.~Seleznev(2024)]{fominov2024}
Ya. V.~Fominov G.~S.~Seleznev.
\newblock Influence of capacitance and thermal fluctuations on the josephson
  diode effect in asymmetric higher-harmonic squids.
\newblock \emph{Physical Review B}, 110:\penalty0 104508, 2024.

\bibitem[Thompson(1973)]{tompson1973}
E.~D. Thompson.
\newblock Perturbation theory for a resistivity shunted josephson element.
\newblock \emph{Journal of Applied Physics}, 44:\penalty0 5587, 1973.

\bibitem[Nashaat et~al.(2019)Nashaat, Bobkova, Bobkov, Shukrinov, Rahmonov, and
  Sengupta]{nashaat2019electrical}
M~Nashaat, IV~Bobkova, AM~Bobkov, Yu~M Shukrinov, IR~Rahmonov, and K~Sengupta.
\newblock Electrical control of magnetization in
  superconductor/ferromagnet/superconductor junctions on a three-dimensional
  topological insulator.
\newblock \emph{Physical Review B}, 100\penalty0 (5):\penalty0 054506, 2019.

\end{thebibliography}
\bibliographystyle{unsrtnat}

\end{document}